\documentclass[graybox]{svmult}

\usepackage{mathptmx}       
\usepackage{helvet}         
\usepackage{courier}        
\usepackage{type1cm}        
\usepackage{makeidx}         
\usepackage{graphicx}        
\usepackage{multicol}        
\usepackage[bottom]{footmisc}

\usepackage[table]{xcolor}

\usepackage{amssymb}
\newcommand{\hzs}{\rm\,Hz\,s^{-1}}
\newcommand{\msun}{\,M_{\odot}}
\newcommand{\nudot}{\dot{\nu}}
\newcommand{\mdot}{\dot{M}}
\newcommand{\rxte}{{\textit{RXTE}}}
\newcommand{\ergs}{\rm\,erg\,s^{-1}}
\newcommand{\mdota}{\langle\mdot\rangle}

\newcommand\aj{{AJ }}%
\newcommand\araa{{ARA\&A }}%
\newcommand\apj{{ApJ }}%
\newcommand\apjl{{ApJ }}%
\newcommand\apjs{{ApJS }}%
\newcommand\apss{{Ap\&SS }}%
\newcommand\aap{{A\&A }}%
\newcommand\actaa{{Acta~Astron. }}%
\newcommand\aaps{{A\&AS }}%
\newcommand\baas{{BAAS }}%
\newcommand\mnras{{MNRAS }}%
\newcommand\pasj{{PASJ }}%
\newcommand\sovast{{Soviet~Ast. }}%
\newcommand\nat{{Nature }}%
\newcommand\iaucirc{{IAU~Circ. }}%
\newcommand\physrep{{Phys.~Rep. }}%
\newcommand\na{{New~Astronomy }}%
\newcommand\nar{{New~Astronomy Reviews}}%
\makeindex             

\begin{document}

\title*{Accreting Millisecond X-Ray Pulsars}
\author{A. Patruno, A. L. Watts}
\institute{A. Patruno \at Leiden Observatory, Huygens Laboratory, Leiden University, Neils Bohrweg 2, 2333 CA, Leiden, The Netherlands;  \email{patruno@strw.leidenuniv.nl}\\ A.L. Watts \at Astronomical Institute ``Anton Pannekoek'', University of Amsterdam, Science Park 904, 1098 XH, Amsterdam, The Netherlands;\email{a.l.watts@uva.nl}}

%
%
\maketitle

\abstract{Accreting Millisecond X-Ray Pulsars (AMXPs) are astrophysical
laboratories without parallel in the study of extreme physics. In this
chapter we review the past fifteen years of discoveries in the field. We
summarize the observations of the fifteen known AMXPs, with a
particular emphasis on the multi-wavelength observations that have
been carried out since the discovery of the first AMXP in 1998.  We
review accretion torque theory, the pulse formation process, and how
AMXP observations have changed our view on the interaction of plasma
and magnetic fields in strong gravity. We also explain how the AMXPs
have deepened our understanding of the thermonuclear burst process, in
particular the phenomenon of burst oscillations.  We conclude with a
discussion of the open problems that remain to be addressed in the
future.}

\section{Introduction}
\label{sec:1}

Neutron stars (NSs), amongst the most extreme astrophysical objects in
the Universe, allow us to study physics in regimes that cannot be
accessed by terrestrial laboratories.  They play a key role in the
study of fundamental problems including the equation of state (EoS) of
ultra-dense matter, the production of gravitational waves, dense
matter superfluidity and superconductivity, and the generation and
evolution of ultra-strong magnetic fields.  Since the discovery of NSs
as radio pulsars in 1967~\cite{hew67}, many different classes have been discovered
including more than ${\sim}130$ NSs in low mass X-ray binaries (LMXBs).
In LMXBs the NS accretes matter from a non-collapsed stellar companion
(with mass $M\lesssim 1\,M_{\odot}$) via an accretion disk.  This
chapter focuses on a subgroup of the LMXBs, the accreting millisecond
X-ray pulsars (AMXPs).  In the AMXPs, the gas stripped from the
companion is channeled out of the accretion disk and onto the magnetic
poles of the rotating neutron star, giving rise to X-ray pulsations at
the spin frequency.  We will explore the details of this process, why
it is so rare (the AMXPs are a small class), the physics of the
disk-magnetosphere interaction, and how AMXPs can be used to probe
extreme physics.

Immediately after the discovery of the first millisecond radio pulsar
in 1982 \cite{bac82}, LMXBs were identified as possible incubators for
millisecond pulsars. It was suggested that LMXBs might be responsible
for the conversion of slow NSs with high magnetic field
($B{\sim}10^{12}$~G), into a rapidly spinning objects with a relatively
weak magnetic field ($B{\sim}10^8$~G).  Two independent papers published
in 1982 \cite{alp82, rad82} proposed transfer of angular momentum
through accretion as the mechanism responsible for the spin-up of
pulsars. This is known as the \emph{recycling scenario} (see
\cite{sri10} for an excellent review). This name originates from the
fact that radio pulsars switch off their pulsed radio emission after
entering the so-called ``pulsar graveyard''.  If this happens while
the NS is in a binary with a non-collapsed low or intermediate mass
stellar companion, binary evolution of the system~\cite{bha91, tau06}
can bring the companion into Roche lobe contact and trigger a
prolonged epoch of mass transfer from the companion (donor) towards
the NS (accretor). The mass is transferred with large specific angular
momentum and the NS is spun-up by the resulting accretion
torques. Once the mass transfer episode terminates, the NS might
eventually switch on again as a ``recycled'' millisecond radio pulsar.

The first AMXP (SAX J1808.4--3658), found in 1998 with the
\textit{Rossi X-ray Timing Explorer} (\textit{RXTE})
\cite{wij98} provided a beautiful confirmation of the recycling
scenario. Fourteen more AMXPs have since been found, with spin
frequencies from 182 to 599 Hz.  Another important milestone came with
the discovery (in 2007) of a binary radio millisecond pulsar (PSR
J1023+0038) for which archival optical observations, taken ${\sim}$ 7
years before the radio pulsar discovery, showed evidence for an
accretion disk\footnote{a new transition from a radio pulsar to a LMXB
  has happened and is currently ongoing at the moment of writing this
  review~\cite{sta13,pat13b}}~\cite{arc09}. This is the first NS
observed to have switched on as a radio pulsar after being an X-ray
binary. A final confirmation that indeed AMXPs and radio pulsars are
related has recently come with the discovery of the system IGR J18245--2452
which has shown both an AMXP and a radio millisecond pulsar phase (see
\cite{pap13a} and Section~\ref{sec:18245} for a detailed discussion).

The \rxte\, observatory has played an extraordinary role by
discovering many systems of this kind and by collecting extensive data
records of each outburst detected during its fifteen year
lifetime. The excellent timing capabilities of \textit{RXTE} have
brought new means to study NSs with coherent X-ray timing, and helped
to constrain the long term properties of many AMXPs over a baseline of
more than a decade.  Observation of the orbital Doppler shift of the
AMXP pulse frequency contains information on the orbital parameters of
the binary and their evolution in time. Binary evolution has benefited
from the study of AMXPs \cite{bil02, nel03, del03} which are now known to
include ultra-compact systems (orbital period $P_b\lesssim 80$ min)
with white dwarf companions, compact systems ($P_b\simeq1.5-3$ hr) with
brown dwarf donors and wider systems ($P_b\simeq3.5-20$ hr) with main
sequence stars. Other X-ray and gamma ray space missions like
\textit{XMM-Newton}, \textit{INTEGRAL}, \textit{Chandra},
\textit{Swift} and \textit{HETE} have also played an important role in
discovering and understanding the spectral and timing properties of
these objects. Multiwavelength observations covering radio, infrared,
optical and UV wavelengths have also illuminated different aspects of
these fascinating systems. Several optical and infrared counterparts
have been identified with ground based observations and in some cases
have led to the discovery of the spectral type of the donor, while
radio and infrared observations have revealed the possible presence of
jets.

This chapter is structured as follows:  

\begin{description}
\setitemindent{largelabel}
\item{\textbf{Section 2.}}  An overview of the AMXP family.
\item{\textbf{Section 3.}}  Observations of individual AMXPs
\item{\textbf{Section 4.}}  Coherent timing analysis and accretion torques.  
\item{\textbf{Section 5.}} X-ray pulse properties, their formation and use.
\item{\textbf{Section 6.}} Long-term evolution of spins and orbital parameters.  
  \item{\textbf{Section 7.}}  Thermonuclear bursts and burst oscillations. 
\item{\textbf{Section 8.}}  Aperiodic  phenomena including kilohertz QPOs.
\item{\textbf{Section 9.}} Future developments and open questions.

\end{description}

\section{The Accreting Millisecond X-ray Pulsar Family}
\label{sec:2}

An AMXP is an accretion powered X-ray pulsar spinning at frequencies
$\nu\ge 100$ Hz bound in a binary system with a donor companion of
mass $M\lesssim 1\msun$ and with weak surface magnetic fields
($B{\sim}10^{8-9}$ G). All known AMXPs have donor stars that transfer mass
via Roche lobe overflow (RLOF). This definition excludes binary
millisecond pulsars, which are detached systems with a pulsar powered
by its rotational energy, X-ray pulsars in high mass X-ray binaries,
symbiotic X-ray binaries and the slow X-ray pulsars in LMXBs with
$\nu<100$ Hz.  Several LMXBs show millisecond oscillations during
thermonuclear bursts that ignite on their surface. These nuclear
powered X-ray pulsars (NXPs), or burst oscillation sources, are not
AMXPs, since they are powered by nuclear burning rather than channeled
accretion (although note that some AMXPs also show burst
oscillations).

Table~\ref{tab:lmxbs} reports the main characteristics of the 23 known
pulsating NSs in LMXBs: 15 AMXPs, 3 slow accreting X-ray pulsars in
LMXBs, 1 slow accreting X-ray pulsar in an intermediate mass X-ray
binary (IMXB), 3 symbiotic X-ray binaries accreting wind
from a K/M giant companion (compatible with being of low mass type)
and 1 mildly recycled accreting pulsar recently discovered in the
globular cluster Terzan 5. This latter source is an 11 Hz X-ray pulsar
with a magnetic field of $10^9-10^{10}$ G and is of particular
interest because it might be the only known accreting pulsar in the
process of becoming an AMXP on a short timescale~\cite{pat12b}. 
The other four slow pulsars are very different systems, in the sense
that they have most likely followed a completely different
evolutionary history, have strong magnetic fields ($B{\sim}10^{12}$~G)
and may never reach millisecond periods. Symbiotic X-ray binaries are
also different from the AMXPs since they are wide binaries
($P_b\gtrsim50$ days) and their NSs have extremely long spin periods
caused by a prolonged phase of spin-down during a wind accretion
process.

All of the AMXPs are transient systems, with accretion disks that run
through cycles of outburst and quiescence. X-ray pulsations have never
been observed in quiescence, although the upper limits are
unconstraining due to the poor photon flux. The shortest outburst
recurrence time is one month for the globular cluster source NGC 6440
X-2, with an outburst duration of less than 4-5 days, whereas the
longest outburst, from HETE J1900.1-2455, has so far lasted for almost
7 years. The small group of AMXPs therefore form a rather
heterogeneous class of objects.  There are, however, some common
features shared by all AMXPs:
\begin{svgraybox}
\begin{itemize}
\item Outburst luminosities are usually faint, suggesting low time averaged
mass accretion rates (Section~\ref{sec:3}). 
\item Spin frequencies appear to be uniformly distributed with an abrupt cutoff at about ${\sim}700$ Hz (Section~\ref{sec:6}). 
\item Ultra-compact binaries are rather common, comprising about 40\% of the
total AMXP population. 
\item Very small donors are preferred, with masses almost always below 0.2$\msun$.
\item The orbital periods are always relatively short, with $P_b<1$~day. Therefore the known AMXPs are probably \textbf{not} the progenitors of wide orbit binary millisecond radio pulsars.
\end{itemize}
\end{svgraybox}

\begin{table}
\caption{Accreting X-ray Pulsars in Low Mass X-ray Binaries}
\scriptsize
\begin{center}
\rowcolors{4}{gray!10}{}
\begin{tabular}{lllllll}
\hline
\hline
Source & $\nu_{s}$ & $P_{b}$ & $f_{x}$  & $M_{c,min}$  & Companion Type & Ref.\\
 & (Hz) & (hr) & ($M_{\odot}$) & ($M_{\odot}$) &   \\
\hline
\textbf{Accreting Millisecond Pulsars}\\
\hline
SAX J1808.4--3658 & 401 &  2.01 & $3.8\times 10^{-5}$ & 0.043  & BD &  \cite{pat12,har08,har09, pap09, cac09}\\
XTE J1751--305 & 435 &  0.71 & $1.3\times 10^{-6}$ & 0.014  & He WD &   \cite{mar02, pap08}\\
\smallskip
XTE J0929--314  & 185  & 0.73 & $2.9\times 10^{-7}$ & 0.0083  & C/O WD  & \cite{rem02b, gal02, nel06}\\
XTE J807--294  & 190  & 0.67 & $1.5\times 10^{-7}$ & 0.0066  & C/O WD  & \cite{mar03b, kir04, cho08, rig08, pat10} \\
XTE J1814--338  & 314  & 4.27 & $2.0\times 10^{-3}$ & 0.17 & MS  & \cite{mar03a, pap07}\\
\smallskip
IGR J00291+5934  & 599  & 2.46 & $2.8\times 10^{-5}$ & 0.039  &  BD & \cite{gal05, fal05b, bur07, pat10b, har11, pap11}\\
HETE J1900.1--2455  & 377  &   1.39 & $2.0\times 10^{-6}$ & 0.016  & BD  & \cite{kaa06}\\
Swift J1756.9--2508  & 182  &  0.91 &  $1.6\times 10^{-7}$ & 0.007  & He WD &  \cite{kri07, pat10c}\\
\smallskip
Aql X--1     & 550 &  18.95 & N/A & $0.6^{a}$ & MS & \cite{cas08, wel00}\\
SAX J1748.9--2021  & 442 & 8.77 &  $4.8\times 10^{-4}$ & 0.1  & MS/SubG ?& \cite{pat09, alt08}\\
NGC6440 X--2 & 206 & 0.95 & $1.6\times 10^{-7}$ & 0.0067 & He WD & \cite{alt10}\\
\smallskip
IGR J17511--3057 & 245 &  3.47 & $1.1\times 10^{-3}$ & 0.13  & MS & \cite{mar09b, rig11c}\\
Swift J1749.4--2807 & 518 & 8.82 & $5.5\times 10^{-2}$ & 0.59 & MS & \cite{alt11}\\
IGR J17498--2921 & 401 & 3.84 & $2.0\times10^{-3}$ & 0.17 & MS & \cite{pap11b}\\
IGR J18245--2452 & 254 & 11.03 & $2.3\times10^{-3}$ & 0.17 & MS & \cite{pap13a}\\
\hline
\textbf{Mildly Recycled X-ray Pulsar}&&&&&\\ 
\hline
IGR J17480-2466 & 11 &  21.27 & $2.1\times10^{-2}$ & 0.4 & SubG &  \cite{str10b, papitto, cav11}\\
\hline
\textbf{Slow Pulsars in LMXBs \& IMXBs}\\
\hline
2A 1822-371 & 1.7  & 5.57 & $2\times10^{-2}$& $0.33$ & ? & \cite{jon01}\\
\smallskip
4U 1626-67 & 0.13  & 0.69 & $1.3\times10^{-6}$& 0.06 & WD or He star  & \cite{bil97, lev88, cob02} \\
GRO 1744-28 & 2.14 & 282.24 & $1.3\times10^{-4}$& $<0.3$ & Giant & \cite{bil97, cui97, zha96}\\
Her X-1 & 0.81 & 40.80 &  0.85 & ${\sim}2^{a}$ & MS & \cite{bil97, cob02}\\
\hline
\textbf{Symbiotic X-ray Binaries}\\
\hline
GX 1+4 &$6.3\times10^{-3}$ &  1161 days & & & M5 Giant& \cite{cui97, hin06, pos10}\\
4U 1954+31 & $5.5\times10^{-5}$ &  &&& M4 Giant &\cite{cor08, pos10}\\
\smallskip
IGR J16358--4724$^b$ & $1.7\times10^{-4}$ &  &&& K/M Giant &\cite{cor08, pos10}\\
\hline
\end{tabular}\\
\end{center}
$\nu_{s}$ is the spin frequency, $P_{b}$ the orbital period, $f_{x}$
is the X-ray mass function, $M_{c,min}$ is the minimum companion mass
for an assumed NS mass of 1.4$\msun$. 
The companion types are: WD = White Dwarf, BD= Brown Dwarf, MS = Main Sequence, SubG = Sub-Giant, He Core = Helium Star.\newline
$^{a}$ The donor mass is inferred from photometric data
and does represent the most likely mass.\newline
$^{b}$ Binary with parameters that are still compatible with an intermediate/high mass donor.
\label{tab:lmxbs}
\end{table}

\subsection{Intermittency}
\label{intermittency}

Until 2007 it was believed that AMXPs showed X-ray pulsations
throughout outbursts.  The seventh AMXP (HETE J1900.1-2455), however,
showed an unexpected new behaviour \cite{gal07}.  For the first
$\approx20$ days of its outburst (which started in 2005) it showed
typical AMXP pulsations. The pulsations then became intermittent,
appearing and disappearing on different timescales for the next
$\approx2.5$ years. Pulsation disappeared and never reappeared again
after MJD 54,499~\cite{pat12c}, with the most stringent upper limits
of $0.05\%$ rms on the fractional pulse amplitude (i.e., a sinusoidal
fractional amplitude of 0.07\%, see Eq.~\ref{fa}).

This discovery was exciting, since it may help to bridge the gap
between non-pulsating LMXBs and AMXPs. It also became immediately
clear that other LMXBs might show sporadic episodes of pulsations
during their outbursts. This has now been found to be the case for two
other sources: Aql X--1 and SAX J1748.9--2021.  Aql X--1 is perhaps
the most striking case recorded so far: coherent pulsations were
discovered in 1998 \textit{RXTE} archival data and appeared in only
one 120s data segment out of a total exposure time of 1.5 Ms from more
than 10 years of observations.  The extremely short pulse episode has
raised discussions about whether it really originated from magnetic
channeled accretion, and whether Aql X-1 can truly be considered an
AMXP. However, the high coherence of the signal leaves little doubt
about the presence of pulsations, and the accretion-powered origin
appears the most promising explanation \cite{cas08}. The case of SAX
J1748.9--2021 is slightly different: pulsations were detected
sporadically in several data segments and in three (2001, 2005 and
2009-2010 \cite{alt08, pat09, pat10e}) out of four outbursts observed
(the first being in 1998).
It is unclear why these three systems show pulsations intermittently.
In HETE J1900.1--2455, an increase in pulse fractional amplitude was
reported approximately in coincidence with the occurrence of Type I
X-ray bursts~\cite{gal07}, followed by a steady decrease. On other
occasions the pulsations appeared a few hours before or after a burst,
indicating that pulsations might be linked somehow with some yet to be
identified property of the NS envelope. In SAX J1748.9--2021 the
pulsed amplitudes showed some abrupt changes in amplitude and/or phase
in coincidence with about 30\% of the observed Type I X-ray bursts
\cite{alt08, pat09}. The pulsations, however, displayed a more diverse
behaviour than in HETE J1900.1--2455, without the typical steady
decrease of fractional amplitudes. A period of global surface activity
during which both Type I bursts and pulsations are produced might be
at the origin of this link~\cite{pat09}. The single pulsating episode
of Aql X-1 had instead no clear connection with Type I bursts, even
though Aql X-1 is a bursting LMXB.

The spin frequency derivative of HETE J1900.1--2455 was measured over
a baseline of 2.5 years, an unprecedented long baseline for an AMXP
(whose outbursts last usually less than 100 days). This has indirectly
provided hints on the physical origin of such a period of global
surface activity in intermittent sources. The spin frequency
derivative exhibited an exponential decay in time that was interpreted
as evidence of the screening of the NS magnetic
field~\cite{pat12c}. It was proposed that intermittency originates
because the magnetic field strength drops by almost three orders of
magnitude on a timescale of few hundred days so that the disk cannot
be truncated and only a (very) shallow layer of gas can be channeled
to form (weak) pulsations (see also Section~\ref{notpulsate} for a
discussion of the model).

One feature that intermittent pulsars share is that the long term
average mass accretion rate $\langle\mdot\rangle$ is higher than for
the persistent AMXPs and smaller than for the bright non-pulsating
systems (like Sco X-1 and the other Z sources). Calculating the
precise value of $\langle\mdot\rangle$ is rather difficult, since it
depends on poorly constrained parameters in most LMXBs, like the
distance $d$, the X-ray to bolometric flux conversion and the
recurrence time of the outburst. However, even considering these
caveats, it seems clear that at least the brightest systems have never
showed pulsations. Whether it is the high mass accretion rate that
determines the lack of pulsations or some other properties shared by
the brightest systems is still unclear.  See Section~\ref{sec:6} for
an extended discussion on the mechanism that might prevent the
formation of pulsations in LMXBs.

\section{Observations of the AMXPs}\label{sec:3}

In this section we discuss the main characteristics of each AMXP, including
number of outbursts observed, luminosity, distance, 
orbital parameters and (at the end of each subsection) radio/IR/optical counterparts.  We devote
Section~\ref{sec:4} and~\ref{sec:6} to the discussion of the results
on the pulsar rotational parameters and their secular evolution,
respectively. We briefly mention results
relating to X-ray bursts, burst oscillations and aperiodic variability
but refer to Section~\ref{sec:7} and~\ref{sec:8} for a general
discussion of these topics. The astrometric position of each AMXP is given
in Table~\ref{tab:pos}.

\begin{table}
\caption{Astrometric Position of AMXPs}
\scriptsize
\begin{center}
\rowcolors{5}{gray!10}{}
\begin{tabular}{llllll}
\hline
\hline
Source & Right Ascension &  Declination & Error & Observatory & Reference\\
       & [HH:MM:SS] & [DD:MM:SS]\\
       &   (J2000)  &  (J2000)& (90\% c.l.)\\
\hline
SAX J1808.4--3658 & 18:08:27.62 & -36:58:43.3 & $0''.15$ & 6.5 m Baade (Magellan I) & \cite{har08} \\
XTE J1751--305 & 17:51:13.49(5) & -30:37:23.4(6) &  & Chandra  & \cite{mar02}\\
\smallskip
XTE J1807--294 & 18:06:59.80 & -29:24:30 & $0''.6$ & Chandra & \cite{mar03b}\\ 
IGR J00291+5934 & 00:29:03.05(1) & +59:34:18.93(5) &  & Multiple Optical Obs. & \cite{tor08}\\
XTE J1814--338 & 18:13:39.04 & -33:46:22.3 & $0''.2$  & Magellan & \cite{kra05}\\
\smallskip
XTE J0929--314 & 09:29:20.19 & -31:23:03.2 & $0''.1$ & Mt. Canopus & \cite{gil05}\\
Swift J1756.9--2508 & 17:56:57.35 & -25:06:27.8& $3''.5$ & Swift/XRT & \cite{kri07}\\
Aql X--1    & 19:11:16.0245341 & +00:35:05.879384 & $0''.0005$ & e-EVN & \cite{tud13}\\
\smallskip
SAX J1748.9--2021 & 17:48:52.163 & -20:21:32.40 & $0''.6$ & Chandra & \cite{poo02} \\
NGC6440 X-2 & 17:48:52.76(2) & -20:21:24.0(1) & & Chandra & \cite{hei10}\\
Swift J1749.4--2807 & 17:49:31.83 & -28:08:04.7 & $1''.6$  & \textit{Swift}/XRT  & \cite{dav11}\\
\smallskip
IGR J17511--3057 & 17:51:08.66(1) & -30:57:41.0(1) & $0''.6$ & Chandra & \cite{now09}\\
IGR J17498--2921 & 17:49:55.38 &   -29:19:19.7 & $0''.6$  & Chandra & \cite{cha11}\\
HETE J1900.1--2455 &  19:00:08.65 & -24:55:13.7 & $0''.2$ & Palomar & \cite{fox05}\\
IGR J18245--2452 &  18:24:32.51 & -24:52:07.9 & $0''.5$ & ATCA & \cite{pav13}\\
\end{tabular}\label{tab:pos}\\
\end{center}
The errors in parentheses refer to R.A. and DEC separately, whereas
those that appear in the column ``Error'' refer to both coordinates.
\end{table}

\subsection{SAX J1808.4--3658}\label{sec:1808}

The source SAX J1808.4-3658 was discovered with the X-ray satellite
\textit{BeppoSAX} during an outburst in September
1996~\cite{int98}. The source showed via three type I X-ray bursts
that the compact object in the binary is a NS. The 2-28 keV peak
luminosity of $3\times 10^{36}\ergs$ \cite{int01} is rather faint and
below the average peak luminosity reached by other LMXBs. X-ray
pulsations were not detected during this outburst, with poorly
constraining upper limits of $20\%$ on the pulsed fraction. Coherent
pulsations at 401 Hz were discovered instead in 1998, during the
second observed outburst~\cite{wij98}, thanks to the better
sensitivity of the Proportional Counter Array (PCA) on
\textit{RXTE}. An orbital modulation of 2.01 hr was detected using
Doppler delays in the coherent timing data \cite{cha98} and it was
suggested that the companion of the pulsar is a heated brown dwarf
with mass of $\approx0.05\msun$ \cite{bil01}.  No thermonuclear X-ray
bursts were observed during the 1998 outburst, but re-analysis of the
1996 data provided marginal evidence for burst oscillations at the NS
spin frequency~\cite{int01}.

\begin{figure}[t]
\sidecaption
\rotatebox{0}{\includegraphics[scale=.31]{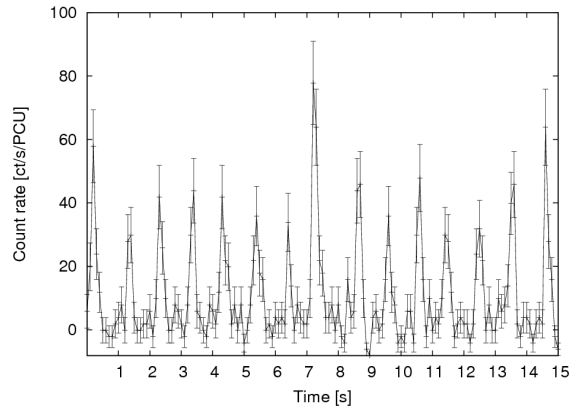}}
\caption{Small portion (15 s) of the 2005 X-ray light-curve of SAX J1808.4--3658, showing the 1 Hz flaring phenomenon during a reflare. The amplitude
of the 1 Hz oscillation reaches 125\% rms (Figure from \cite{pat09c}).}
\label{fig:1hz}       
\end{figure}

SAX J1808.4--3658 went into outburst again in 2000, 2002, 2005,
2008 and 2011, with an approximate recurrence time of about 1.6-3.3
yr, and is the best sampled and studied of all AMXPs. All of
the outbursts showed faint luminosities, with peaks always below
$10^{37}\ergs$, even when considering broad energy bands.  The 2000
outburst was poorly sampled due to solar constraints, and was
observed only during the ``flaring tail'', a peculiar outburst phase
displayed by only a very small number of X-ray sources. The typical
SAX J1808.4-3658 outburst can be split in five phases: a fast rise,
with a steep increase in luminosity lasting only a few days, a
peak, a slow decay stage, a fast decay phase and the flaring tail.
The first four phases are typical of several X-ray binaries and dwarf novae
and can in principle be partially explained with the disk instability
model. The flaring tail instead shows bumps, called ``reflares'', with
a quasi-oscillatory behaviour of a few days and a variation in
luminosity of up to three orders of magnitude on timescales 
${\sim}$ 1-2 days~\cite{wij01,wij03,pat09c}. The flaring tail has no
clear explanation within the disk instability model~\cite{las01}, but
has been observed in all outbursts~\cite{pat09c,gal06}.
Pulsations are still detected during this low luminosity stage, and a strong ${\sim} 1$ Hz oscillation is observed to
modulate the reflares (see
Figure~\ref{fig:1hz}; \cite{vank00, pat09c}).  This has been interpreted
as a possible signature of the onset of a disk instability with radial
perturbation of the magnetospheric boundary (the so-called ``Spruit
and Taam instability''~\cite{spr93, dan10}) close to the onset of
the``propeller stage''~\cite{pat09c}. The 1 Hz modulation, however,
has not appeared in all the flaring tails of the different outbursts,
but only in the years 2000, 2002 and 2005. The presence of the 1 Hz
modulation during the 1998 flaring tail cannot be excluded, however, due to lack
of observations.

The study of the high frequency aperiodic variability of SAX
J1808.4--3658 led to a breakthrough when twin kHz Quasi Periodic
Oscillations (QPOs) at a frequency of approximately 500 and 700 Hz
were discovered for the first time in a source with a well established
spin frequency~\cite{wij03b}. The twin kHz QPOs (Section~\ref{sec:8})
were observed when SAX J1808.4--3658 reached the peak luminosity
during the 2002 outburst, and their frequency separation of $196\pm4$
Hz is consistent with being at half the spin frequency. This suggested
a possible link between spin frequency and kHz QPOs and hence refuted
the beat-frequency model for the formation of kHz QPOs. In this model
the upper kHz QPO at 700 Hz reflects an orbital frequency in the inner
accretion disk, whereas the lower kHz QPO should appear as a beat
frequency between the upper kHz QPO and the spin frequency. The twin
kHz QPOs must be separated by a frequency equal to the spin frequency
and not half its value, as observed.  The upper kHz QPO was observed
during most of the outburst, and another QPO at 410 Hz also appeared
during a few observations~\cite{wij03b}. The origin of this latter QPO
is unclear and it has been suggested~\cite{wij03b} that it might be
related to the side-band phenomenon observed in other
LMXBs~\cite{jon00}. In particular, it was suggested that the 410 Hz
QPO might be a sideband of the pulsation, created by a resonance
occurring at the radius in the disk where the general relativistic
vertical epicyclic frequency matches the spin frequency of the pulsar.

The 3-150 keV X-ray spectrum of the 1998 outburst had a remarkably
stable power law shape with photon index of ${\sim}2$ and a high energy
cutoff at ${\sim}$ 100 keV \cite{gil98}. Further spectral analysis
provided evidence for a two-component model: a blackbody at soft
energies and a hard Comptonization component at higher energies
\cite{gie02}. The blackbody is interpreted as the heated hot spot on
the NS surface, whereas the Comptonization is produced in the
accretion shock created at the bottom of the magnetic field lines as
the plasma abruptly decelerates close to the NS surface.  The presence
of an accretion disk was detected much later at lower energies, with
observations taken in 2008 with the EPIC-pn camera on the
\textit{XMM-Newton} telescope.  Its large sensitivity at soft energies
(down to about 0.5 keV) well below the nominal 2 keV limit of
\textit{RXTE}, allowed the detection of the typical cold accretion
disk signature at a temperature of 0.2 keV \cite{pap09,pat09e}.  The
signature of a fluorescent relativistic iron K$\rm\alpha$ emission
line profile was also found. A similar result was obtained with
combined \textit{XMM-Newton} and \textit{Suzaku} data
\cite{cac09}. Spectral modeling of the iron line constrained the
magnetic field of the pulsar to be ${\sim}3\times10^8$~G at the
poles~\cite{cac09, pap09}. Simultaneous spectral modeling of the inner
disk radius and pulse profile shapes of the 2002 outburst~\cite{ibr09}
lead to a similar constraint of the magnetic field ($B{\sim}10^{8}$~G).

\begin{figure}[t]
\sidecaption
\includegraphics[scale=.35]{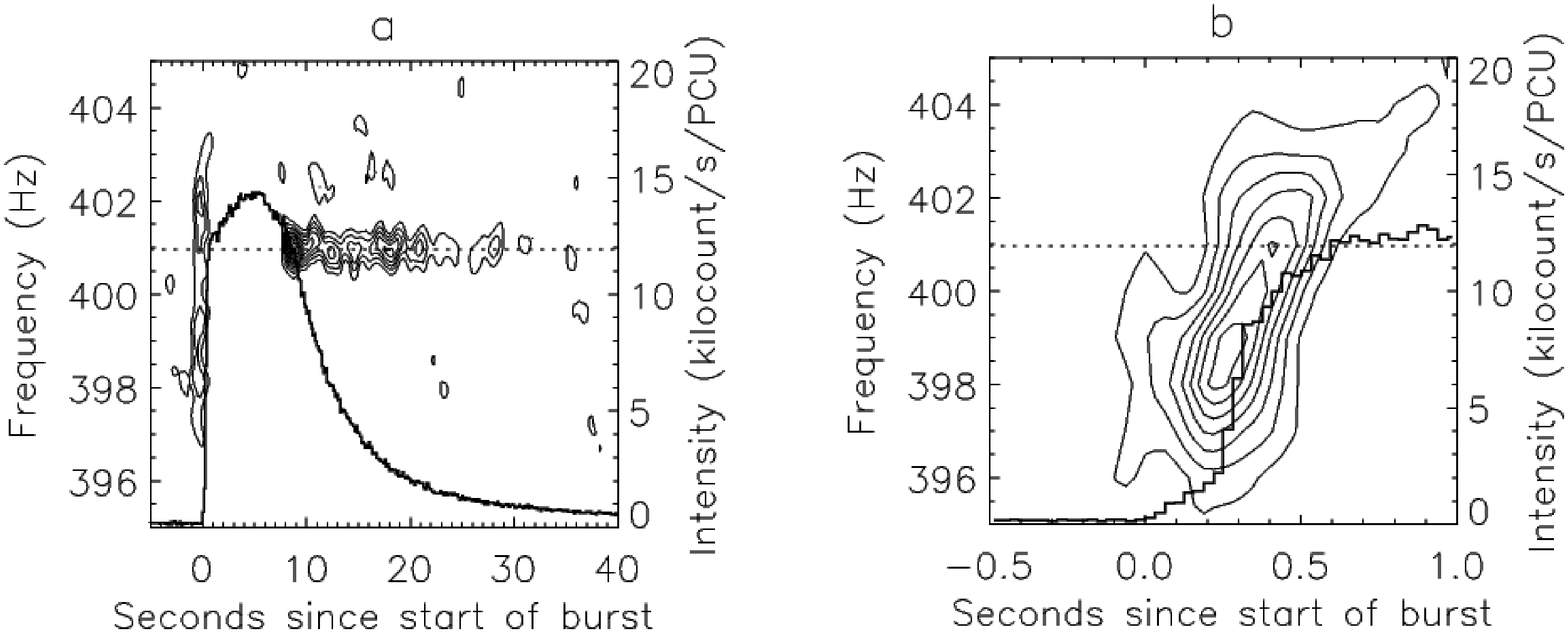}
\caption{\textbf{Left panel:} a typical X-ray burst observed in the
  2-60 keV energy band on 18 October 2002 with a dynamical power
  spectrum of barycentered data superimposed.  The contours refer to
  power levels of 2 s power spectra spaced by 0.25 s. The burst
  oscillations are clearly detected during the rise, overshooting the 
spin frequency of SAX J1808.4-3658 at 401 Hz. The oscillations then 
disappear during the photospheric radius expansion close to the burst
peak and reappear in the burst tail at a frequency consistent with the
401 Hz of the pulsar. \textbf{Right panel:} a zoom in of the burst rise and
peak, with finer power spectra resolution of 0.25 s spaced by 0.03125 s. The
frequency overshoot is clearly visible, with the BO frequency reaching 
${\sim}403$ Hz (Figure from \cite{cha03}).}
\label{fig:4}       
\end{figure}

Thermonuclear bursts were observed in 1996, 2002, 2005, 2008 and 2011
\cite{int01, cha03, gal06}.  Most of the bursts exhibit photospheric
radius expansion (PRE), where the luminosity reaches the Eddington
limit, lifting the photosphere off the surface of the NS until the
flux dies down.  Such bursts can be used as standard
candles~\cite{kuu03}.  The distance estimated using this method is
2.5-3.6 kpc \cite{gal06, int01}. Note that a different lower limit of
3.4 kpc is reported in \cite{gal06}.  This lower limit is based on the
assumption that the long-term mass transfer rate is driven purely by
loss of angular momentum in the binary via emission of gravitational
radiation, which may not be a good approximation (see
Section~\ref{sec:6}).  All bursts observed in the \textit{RXTE} era
have shown burst oscillations (with a possible marginal detection in
one burst in 1996 observed with \textit{BeppoSAX}) with an amplitude
of a few percent rms. SAX J1808.4-3658 provided the first robust
confirmation that burst oscillation frequency was, to within a few Hz,
the spin frequency of the star (\cite{cha03} and Figure \ref{fig:4}).

An optical/IR counterpart (V4584 Sagittarii) was discovered during the
1998 outburst, coincident with the position of SAX
J1808.4-3658~\cite{roc98}. The reported magnitudes of the candidate
were V = 16.6, R = 16.1, I = 15.6, J = 15.0, H = 14.4, K = 13.8, with
an uncertainty of 0.2 mag in VRI and $<0.1$ mag in JHK. The V band was
further monitored \cite{gil99} and a possible sinusoidal modulation at
the 2 hr period of the binary identified, together with a decay in
the luminosity as the outburst progressed in its decay stage. A
multiband optical/IR photometric study of the optical counterpart during the 1998 outburst revealed an optical flux
consistent with an X-ray heated accretion disk and an inclination of
$\rm\,cos\it\, i\rm\,= 0.65^{+0.23}_{-0.33}$ (90\% c.l.)\cite{wan01}.  The IR
observations, however, showed an excess with respect to an accretion
disk plus irradiated donor star model. This excess was transient in nature,
as it was detected only during one observation, the rest being
consistent with the irradiated accretion disk plus donor star model.
Radio observations carried out one week later (April 27, 1998)
with the Australia Telescope Compact Array (ATCA) revealed a transient
radio counterpart with a flux of ${\sim}0.8$ mJy with no
further detection at later epochs during the 1998 outburst or during
quiescence. This observation was interpreted as synchrotron emission
which might in principle constitute a viable explanation also for the
IR transient excess. This suggests that material can be ejected
from the binary via relativistic jets and/or outflows.
A similar IR excess was detected during 2005 June 5~\cite{gre06} and
radio Very Large Array (VLA) observations carried at 4.86 and 8.46 GHz
showed a transient ${\sim}0.4$ mJy flux, again interpreted
as possible synchrotron emission\cite{rup05d}. 

Optical observations in \textit{quiescence} have been performed extensively
for over a decade~\cite{hom01,del08,wan13}. In the first observation,
a clear optical excess was detected, well above the level expected from
residual X-ray irradiation of the donor star that is still present
during quiescence~\cite{hom01}.  A clear modulation at the orbital
period of the binary in IR bands was also identified~\cite{del08,
  wan09, wan13}. To explain the observations, a strong source of extra
irradiation was required. In particular, the initial donor mass
estimate of 0.05$\msun$~\cite{bil01} was revised to 0.07-0.11$\msun$,
to account for the large entropy of the donor in quiescence
\cite{del08}. The optical excess was first interpreted as evidence for
the turning on of an active radio pulsar when accretion halts
\cite{bur03, cam04}.  Whether this is the correct interpretation is
still to be verified, but it appears a plausible explanation
even though no radio pulsations have been detected~\cite{iac10}.

\subsection{XTE J1751--305}

The AMXP XTE J1751--305 was discovered by \textit{RXTE} on April 3,
2002 and coherent pulsations at 435 Hz were observed immediately
\cite{mar02}. The orbital period of 42 minutes makes this AMXP an
ultra-compact binary with a heated He or C/O white dwarf of minimum
donor mass 0.013--0.017$\msun$ \footnote{Hydrogen rich donor stars
  never reach orbital periods below ${\sim}80$ min~\cite{rap82}}
(depending on the NS mass)~\cite{del03}. Archival observations taken
with the All-Sky-Monitor (ASM) aboard \rxte\, showed that a faint
outburst had already occurred in 1998. Three other outbursts were
observed in 2005, 2007 and 2009. The first two had very faint peak
luminosities of about 10\%-20\% the value reached in 2002. The 2005
outburst in particular was observed by \textit{INTEGRAL} \cite{gre05}
and lasted only 2 days, with only one follow-up \textit{RXTE}
observation \cite{swa05}. No high resolution timing data were taken,
and the association of the outburst with XTE J1751--305 remains
dubious given the large source position error of the two X-ray
observatories. On April 5, 2007, XTE J1751--305 was observed with
\textit{RXTE} during routine galactic bulge monitoring
observations~\cite{mar07}. No pulsations were seen because of the dim
flux (10 mCrab in the 2-10 keV) and the short bulge scan exposure
(${\sim}50$ s).  However, a follow-up \textit{Swift} observation made
the identification of the source certain thanks to the high angular
resolution of the XRT telescope~\cite{mar07b}. The outburst had a
fluence of about $10^{-4}\rm\,erg\,cm^{-2}$, comparable to the already
low 2002 outburst fluence of
$2.5\times10^{-3}\rm\,erg\,cm^{-2}$~\cite{mar02}.  In 2009
\textit{INTEGRAL} recorded a new outburst \cite{che09} and
\textit{RXTE} observed it extensively, with pulsations immediately
detected\cite{mar09}. This outburst has a duration and brightness
similar to the 2002 event.

\begin{figure*}[th!]
 \centering 
$\begin{array}{cc}

 \includegraphics[scale=0.2]{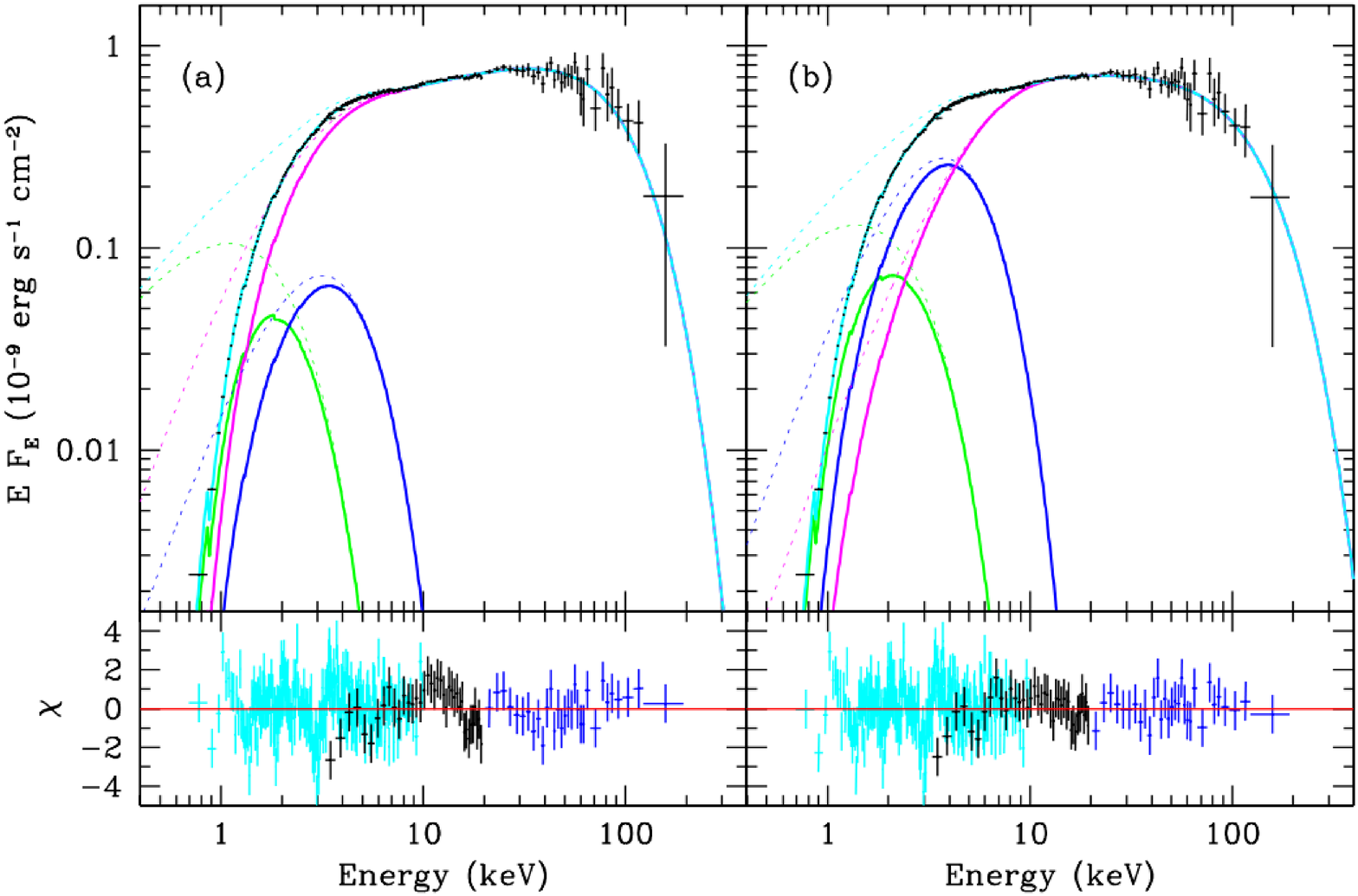}&
   \includegraphics[scale=0.2]{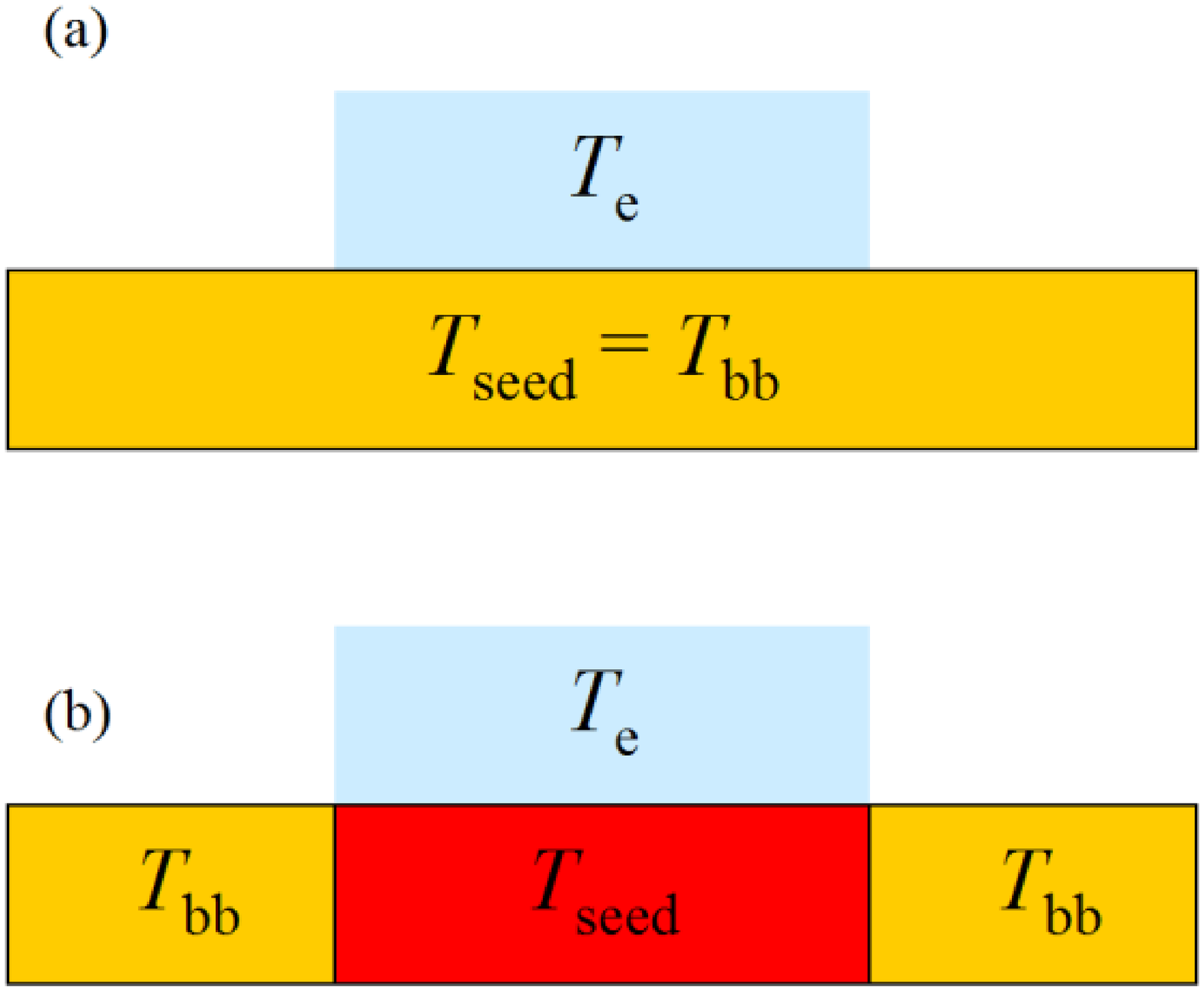}
\end{array}$
 \caption{Broadband X-ray spectrum of XTE J1751--305 obtained by
   combining \textit{XMM-Newton} (light blue points) and \textit{RXTE}
   data (black and dark blue points). Three components are used to
   model the spectrum: a cold blackbody identified with the thermal
   disk emission (green line), a hot blackbody (hot spot; blue line))
   and a hard tail created via thermal Comptonization in a plasma
   (pink line). The left spectrum refers to the emission geometry
   ``(a)'' in which a small temperature gradient is present in the hot
   spot. The right spectrum uses the ``(b)'' model, with a large
   temperature gradient between the hot spot and the remaining
   illuminated NS surface around the accretion shock (image taken from
   \cite{gie05})}
 \label{fig:5}
\end{figure*}

XTE J1751--305 shows no Type I X-ray bursts and its distance
is therefore difficult to determine. The X-ray spectrum taken with
\textit{XMM-Newton} in 2002 showed a power law with spectral index
$1.44\pm0.01$ contributing 83\% of the total 0.5--10.0 keV flux and a
blackbody at temperature $kT=1.05\pm0.01$ keV~\cite{mil03}.  A
broadband spectrum analysis of the 2002 outburst was performed by
combining \textit{RXTE} (PCA and HEXTE) and \textit{XMM-Newton}
(EPIC-pn) data~\cite{gie05}. The analysis revealed the presence of
three components, two blackbodies and a hard component, which were
interpreted as a thermal emission from a cold accretion disk ($kT{\sim}
0.6$ keV), a hotter ($kT{\sim}1$ keV) hot spot on the NS
surface and a thermal Comptonization in plasma of temperature
$kT{\sim}40$ keV (see Fig~\ref{fig:5}). These findings were very similar
to the spectral modeling described above for SAX J1808.4-3658.  Unsuccessful searches for an optical/near IR counterpart were carried out during the 2002 outburst and in quiescence  \cite{jon03}.  However the upper limits are not particularly constraining if one
assumes that the source is close to the Galactic center ($d=8.5$ kpc).

\subsection{XTE J0929--314}

With a spin of 185 Hz and an orbital period of 43.6 min, XTE
J0929--314 was the third AMXP to be discovered, and the second in an
ultra-compact binary.  The source was first detected on April 13, 2002 by the 
\textit{RXTE}/ASM \cite{rem02}.  Coherent pulsations were first detected
in an observation performed on May 2, 2002 \cite{rem02b}. Doppler
modulation of the pulsations revealed the ultra-compact nature of the
binary~\cite{gal02}. The very small mass function
($2.7\times10^{-7}\msun$) implied a tiny donor star of minimum mass of
about 0.01$\msun$. This requires a cold He white dwarf companion with
a possible H-rich envelope that would explain the detection of an
$H_\alpha$ line at $\lambda=6,563~\AA$ \cite{cas02} during the 2002
outburst. The 2002 outburst lasted at least 52 days,
but only 34 days were monitored by the high time resolution
\textit{RXTE}/PCA instrument. The X-ray spectrum of XTE J0929--314,
taken with \textit{Chandra}, showed a power-law plus blackbody similar
to XTE J1751-305~\cite{jue03}.

XTE J0929--314 has been detected in several wavebands: radio, near IR
and optical counterparts were identified during the 2002 outburst.  An
optical counterpart consistent with the position of XTE J0929--314 was
detected in V, B, R and I bands \cite{gre02}. Broad band BVRI
observations at the Mt Canopus 1-m telescope carried in May 2002
revealed a modulation of $\approx 10\%$ in the I band with periodicity
consistent with the orbit of the binary \cite{gil05}. An anomalous
I-band excess was also observed in the first few days of the outburst
and was interpreted as synchrotron radiation, in addition to the
accretion disk/irradiated donor emission. This behavior is similar to
that reported for SAX J1808.4--3658.  The optical
spectrum of the source (in outburst) shows a strong emission line at
4,640$~\AA$ compatible with a C/O or a He/N disc~\cite{nel06}.  Optical
spectroscopy also lead to the conclusion that the most likely donor of
XTE J0929--314 is a C/O white dwarf.  A radio counterpart (during the
outburst) was detected with the VLA at 4.86 GHz
\cite{rup02b}. A source flux of $0.31\pm0.07$ mJy (May 3) and
$0.36\pm0.05$ mJy (May 7) was reported from a position only 0.8
arcseconds from the previously detected optical position
~\cite{gre02}.  The source has also a very faint optical counterpart
($R\,=\,27.2$) in quiescence~\cite{mon05}. PSF $VRI$ photometry in
quiescence~\cite{dav09} also suggested that the donor is irradiated by
a source which emits in excess (by a factor of $\approx8$) of the
quiescent X-ray luminosity of XTE J0929--314
($L_{X}{\sim}4.5\times10^{31}\ergs$).

\subsection{XTE J1807--294}

The fourth AMXP was discovered on February 21, 2003 by \textit{RXTE}
~\cite{mar03b}.  The source is the third ultra-compact AMXP, with the
shortest orbital period known to date, 40.1 min, and a minimum
companion mass of 0.0066$\msun$. Simultaneous \textit{INTEGRAL},
\textit{XMM-Newton} and \textit{RXTE} spectral analysis \cite{fal05a}
constrains the companion mass to be less than 0.022$\msun$, implying
that the donor is likely a C/O white dwarf rather than a He white
dwarf, which would require unlikely a-priori low inclinations.  With
the exception of the quasi-persistent intermittent AMXP HETE
J1900.1-2455, XTE J1807--294 had the longest AMXP outburst observed so
far: $\approx$ 120 days, of which $\approx$ 109 days had sufficiently
high flux to allow the detection of pulsations.  Twin kHz QPOs were
detected in several observations and the separation $\Delta\nu$ of the
kHz QPOs is consistent with the spin frequency of the
AMXP~\cite{lin05}.  No counterparts have been reported at any
wavelength during either outburst or quiescence. A candidate optical
counterpart detection was reported in 2009 \cite{dav09} with $V {\sim}
22.1$ and $R {\sim} 21.4$ along with a tentative detection in the $J$
band. The counterpart falls within the $0''.6$ \textit{Chandra} (90\%
confidence level) error circle. No variability has been observed, with
upper limits of 0.1 mag.

\subsection{XTE J1814--338}

The fifth AMXP was found during routine Galactic center observations
by the \textit{RXTE}/PCA on June 5, 2003 \cite{mar03a}.  The pulse
frequency is 314 Hz and the orbital period 4.3 hr, giving a minimum
companion mass of 0.17$\msun$.  The outburst lasted for $\approx$ 50
days and pulsations were detectable for $\approx$ 45 days. The source
showed 28 Type I X-ray bursts and \textit{RXTE} detected burst
oscillations at the spin frequency in all bursts. The 28th burst
showed signs of PRE, allowing a determination of the source distance
of $\approx 8$ kpc.  The X-ray position of XTE J1814--338 is
consistent with the \textit{EXOSAT} source EXMS B1810-337, detected in
September 1984 \cite{wij03c}. If this identification is correct, then
XTE J1814--338 has a recurrence time of ${\sim}20$ yr.

The optical counterpart was first tentatively identified using BVRI
data taken with the Magellan 6.5 m telescope~\cite{kra03}.  An
observation of H and He emission lines, with a double-peaked
$H_\alpha$ line typical of interacting binaries, was reported soon
after the discovery of the binary \cite{ste03}.  The counterpart was
studied in more detail and finally unambiguously
identified~\cite{kra05}. The B, V and R counterparts during the 2003
outburst had magnitudes between 18 and 19 mag, whereas the I band was
observed with an excess flux reaching about 17.4 mag. This excess is
again similar to that reported for SAX J1808.4--3658 and XTE
J0929--314 and suggests the presence of jets and/or outflows. 
A faint object with $V{\sim}
23.3$ and $R{\sim}22.5$ was identified in ESO Very Large Telescope (VLT)
archival data taken in May 20--21, 2004 \cite{dav09}.  
A multiband Very Large Telescope (VLT) campaign carried in 2009 
(during quiescence) shows an irradiated companion star that requires an energy source
compatible with the spin-down luminosity of a millisecond pulsar \cite{bag13}.
This proves further evidence that AMXPs might turn on as radio pulsars when in quiescence. 

\subsection{IGR J00291+5934}\label{00291}

IGR J00291+5934 is the fastest AMXP known, with a spin of 599 Hz. The
source was discovered by \textit{INTEGRAL} on 2004 December 2
\cite{eck04} and prolonged \textit{RXTE} observations lead to the
discovery of pulsations \cite{mar04} and an orbital period of 2.46 hr,
with a minimum donor mass of about 0.04$\msun$ \cite{mar04b}.  The 2004
outburst had a duration of $\approx$ 14 days from the \textit{INTEGRAL}
discovery, with a smooth flux decay.  On 2008 August 13 the source started a new outburst \cite{cha08b}
and was observed for about a week before reaching an X-ray
flux level below the sensitivity limit of \textit{RXTE}.  The peak
flux observed was about half the value reached in 2004. Follow up
\textit{Swift}/XRT observations on August 21 revealed a flux
level below $10^{-11}-10^{-12}\ergs$ and an \textit{XMM-Newton}
observation on August 25 provided a 2--10 keV flux of
$(1.4\pm0.3)\times10^{-14}\ergs$~\cite{lew10}, compatible with the $10^{-13}\ergs$
(0.5--10 keV) quiescent flux level observed two months after the end of
the 2004 outburst \cite{jon08}.  On 2008 September 21, the source re-brightened again, showing another
outburst lasting for about 14 days. The fluence of the second and
third outbursts were very similar and it is unclear whether the two
2008 outbursts were distinct \cite{har11} or part of the same outburst.  An analysis of archival
\textit{RXTE}/ASM data revealed possible outbursts in 1998
November and 2001 September, suggesting a recurrence time of 3--4 yr
\cite{rem04}.

The pulsar has shown no thermonuclear bursts despite very similar
orbital parameters to SAX J1808.4--3658. The donor star of IGR
J00291+5934 is also probably a X-ray heated brown dwarf as in SAX
J1808.4--3658 \cite{gal05}. Broadband spectral observations performed
during the 2004 outburst with \textit{INTEGRAL} and \textit{RXTE} were
well fitted with a thermal Comptonization model with an electron
temperature of 50 keV and Thomson optical depth $\tau_{\rm T}{\sim}1$ in
a slab geometry \cite{fal05b}. The \textit{RXTE} data analysis of the
first 12 days of the outburst revealed a power law plus blackbody
(with $kT{\sim}1\rm\,keV$) interpreted as emission from the hot spot on the
NS surface.  Further simultaneous \textit{RXTE} and
\textit{Chandra}/HETGS (High Energy Transmission Grating Spectrometer)
spectral observations collected towards the end of the 2004 outburst
revealed a power law plus a cold black body at 0.42 keV, interpreted
as the cold accretion disk, with no signature of the hot spot
blackbody~\cite{pai05}. This is consistent with the lack of detected
pulsations towards the end of the outburst \cite{gal05}.


A tentative optical counterpart was identified during the 2004
outburst~\cite{fox04}, with an R magnitude of 17.4. Follow up spectral
measurements with the 4.2-m William Herschel Telescope on La Palma
strengthened the association with the identification of a weak HeII
and H$_\alpha$ line~\cite{roe04}.  An optical/NIR photometric study in
quiescence was then performed in 2005 with the 3.6-m Telescopio
Nazionale Galileo (TNG) \cite{dav07}. The VRIJH counterparts of IGR
J00291+5934 were detected, with a strong upper limit in the
K-band. The optical light curve shows variability consistent with the
2.46 hr orbital period. The radio to X-ray Spectral Energy
Distributions (SEDs) revealed a blue component indicative of an
irradiated disc \cite{lew10}. The SED contained also a transient NIR
excess similar to that found in SAX J1808.4--3658, XTE J0929--314 and
XTE J1814--338. An $H_\alpha$ line was observed in the optical
spectrum of the 2004 and 2008 outbursts. Evidence was also found for
an $I$ band excess during quiescence in an observation performed with
the 4.2-m William Herschel Telescope (WHT) on 2006 September 13 and 14
\cite{jon08}.

Radio observations taken on 2004 December 4 with the Ryle Telescope in
Cambridge at 15 GHz resulted in a detection at 1.1 mJy \cite{poo04}.
Follow up observations taken one day later gave a non detection,
with upper limits of 0.6 mJy at 15 GHz \cite{fen04}. The Westerbork
Synthesis Radio Telescope (WSRT) observed IGR J00291+5934 between 2004
December 6 and 7, and a radio counterpart was seen at 5 GHz with a
flux of $0.250\pm0.035$ mJy~\cite{fen04}. Very Large Array (VLA)
observations performed on 2004 December 9 at 4.86 GHz confirmed the
reported radio detection, with a flux of $0.17\pm0.05$ mJy.  The
source was not detected in radio during the 2008 outburst, with upper
limits of 0.16 mJy at 5 GHz in an observation taken between August 15
and 16 \cite{lin08}.

\subsection{HETE J1900.1--2455}

The \textit{High Energy Transient Explorer} (HETE--2) discovered HETE
J1900.1--2455 on 2005 June 14 via an X-ray burst \cite{vand05}.
Follow up observations taken on June 16 by \textit{RXTE}/PCA showed
pulsations at 377 Hz \cite{mor05}.  A source distance ${\sim}4$--$5$~kpc
(assuming a typical NS mass of 1.4$\msun$ and a He rich NS atmosphere)
was determined using the first type I X-ray burst detected with
HETE--2 \cite{kaw05, suz07}, which showed PRE.  The orbital period was
determined to be 1.4 hr with a minimum companion mass of
0.016$\msun$\cite{kaa06}.  Unlike the other AMXPs with brown dwarf
companions discussed so far, the donor star in HETE J1900.1--2455 is
most likely a brown dwarf which does not require X-ray irradiation or
a non-standard evolution to fill its Roche lobe.  

A single 882.8 Hz QPO was detected in the \textit{RXTE} lightcurve
during a bright flare occurred on MJD 53,559.45~\cite{kaa06}.
Simultaneous \textit{RXTE} and \textit{INTEGRAL} observations showed
that the 2--300 keV spectrum is well fit by a blackbody component
peaking at 0.8 keV plus thermal Comptonization with electron
temperature of 30 keV and optical depth $\tau_T{\sim}2$ for a slab
geometry \cite{fal07}.  The source was detected up to energies of 200
keV and the 0.1--200 keV luminosity was inferred to be
$5\times10^{36}\ergs$.  Although the peak luminosity is not
particularly high compared to other AMXPs, the average mass accretion
rate $\mdota$ of HETE J1900.1--2455 is among the highest observed due
to its prolonged X-ray outburst that has lasted for eight years (and
is still active as of November 2013)~\cite{gal06c, gal08}. This makes
HETE J1900.1--2455 a so called ``quasi-persistent source'', i.e., an
X-ray binary with a prolonged outburst duration of several years.
Despite the long baseline provided by the outburst, pulsations were
\textit{continuously} detected only in the first two months of the
observations although sporadic pulse detections occurred over a
baseline of about 2.5 yrs (Section~\ref{intermittency}).

An optical counterpart was identified with the Robotic Palomar 60-inch
Telescope on June 18~\cite{fox05}. The source had a brightness
$R{\sim}18.4$.  Near IR observations taken with the 1.3~m robotic
PAIRITEL telescope on Mt. Hopkins indicated a counterpart with
$J=17.6\pm0.2$, with no variability between observations taken on June
16 and 18~\cite{ste05}. The counterpart was observed again with the
FLWO 1.2-m telescope at Mt. Hopkins with $V=18.09\pm0.03$,
$R=18.02\pm0.03$ and $I=17.88\pm0.02$~\cite{ste05b}.  A spectral
analysis of the optical counterpart showed a HeII 4,686$~\AA$,
suggesting correct identification of the binary's optical counterpart.
Further optical observations taken with the 8.4-m Large Binocular
Telescope (LBT) on Mt. Graham, Arizona, showed a brightness of
$R=18.64\pm0.02$ and $18.54\pm0.02$ on 2007 June 13, after an initial
decline of $1.7$ mag in optical, and a non-detection in X-rays which
occurred two weeks prior \cite{deg07, tor07}.  Phase resolved optical
photometry and spectroscopy obtained in September 2006 at the 2.3-m
ATT, 3.5-m WIYN and 8.4-m LBT~\cite{ele08}, revealed a sinusoidal
modulation in the R-band, close to the orbital period. However,
further analysis showed that the modulation is probably due to a
super-hump of a precessing accretion disk, and not irradiation of the
donor star. A broad H$_{\alpha}$ emission line in a 10-m Keck I
spectrum taken by the same authors confirmed this interpretation. No
radio counterpart was identified in VLA observations on 2005 June 19
and 24 at 8.46 GHz~\cite{rup05b}.

\subsection{Swift J1756.9--2508}

\begin{figure}[t]
\sidecaption
\includegraphics[scale=.35]{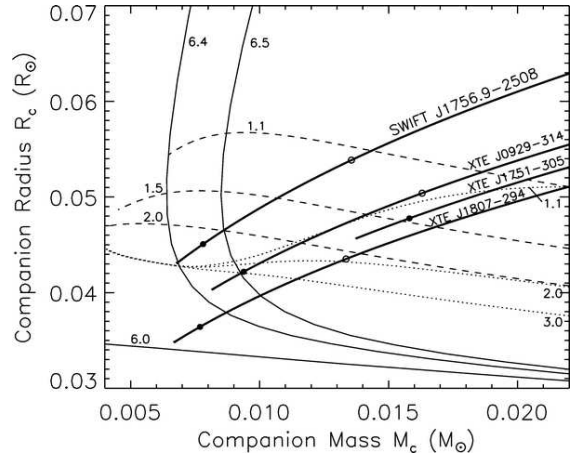}
\caption{Mass-Radius relation (solid line) for the companions of
  the four ultra-compact AMXPs, as determined from the mass function.
Open and filled circles indicate an inclination of $60^{\circ}$
  and $30^{\circ}$, respectively.  Dotted and dashed curves are
  non-irradiated/irradiated white dwarf models, respectively. The
  numbers next to the curves indicate the degeneracy parameter. The
  solid thin curves are pure C white dwarf models and the numbers
  refer to the log of the temperature (in Kelvin). (Figure from~\cite{kri07}) }
\label{fig:m-r}       
\end{figure}

Swift J1756.9--2508 was discovered by the Burst Alert Telescope (BAT)
aboard \textit{Swift} on June 7, 2007~\cite{kri07}.  High time
resolution data from the \textit{RXTE}/PCA, showed that the source was
a 182 Hz AMXP ~\cite{mar07c}. This was the fourth ultra-compact AMXP
discovered, with an orbital period of 54.7 min. The minimum donor star
mass, given by the X-ray mass function, is 0.0067$\msun$.  It has been
proposed that Swift J1756.9--2508 harbors a He rich white dwarf which
is irradiated by X-ray flux from the accretor (see
Figure~\ref{fig:m-r} and \cite{kri07}).  The spectrum of the source,
taken with \textit{Swift}/XRT, BAT and \textit{RXTE}/PCA, is well fit
by a single absorbed power law with spectral index
$\Gamma{\sim}2$~\cite{kri07}. The 2007 outburst length was 13 days,
with no prior outbursts seen in archival data. A second outburst was
detected on 2009 July 8 by \textit{Swift}/BAT and
\textit{RXTE}/PCA~\cite{pat09d}.  This second outburst showed a very
similar length and shape as the 2007 outburst. No thermonuclear bursts
were observed. There is evidence for a fluorescent Fe line in the
X-ray spectrum, although the poor spectral resolution of \textit{RXTE}
prevented to constrain the physical parameters of the
system~\cite{pat10c}. A possible NIR $K_s$ counterpart was
identified~\cite{bur07b} by inspecting the variability of 70 objects
within the 3.5 arcsec error circle of the~\textit{Swift}/XRT
position. The candidate counterpart exhibited variability of about 1.3
mag in two observations taken close to the end of the outburst (2007
June 19) and during quiescence (2007 July 1).  No radio counterpart
has been identified~\cite{pos07, hes07} and Chandra observations taken
on 2007 July 7 detected one single photon in a 11 ks long
observations~\cite{pap07b}.

\subsection{Aql X--1}

Aql X-1 is perhaps the most peculiar AMXP since it has showed
pulsations at 550.27 Hz in just 120 s of X-ray data \cite{cas08}: the
short baseline means that no coherent timing solution can be
obtained. As a consequence, no orbital modulation or mass function has
yet been measured with X-rays. The source shows thermonuclear bursts
with burst oscillations whose asymptotic frequency is $\approx 0.5$ Hz
smaller than the coherent accretion powered pulsations observed in
1998~\cite{zha98}. The recurrence time of the outbursts is about 1 yr
and they have been extensively monitored at different
wavelengths. Twin kHz QPOs were observed at a frequency of ${\sim}800$
and ${\sim}1080$ Hz~\cite{bar08}, so that their frequency separation is
consistent with being equal to half the NS spin frequency.  The
optical counterpart of Aql X-1, named V1333 Aql, was identified in
1978 \cite{tho78} as a star with $B{\sim}17$ mag (in outburst) and
$B>20$ mag (in quiescence).  The spectral type of the source was
identified as a K6-M0 star \cite{che99}. An orbital modulation at a
period of 18.95 hr was reported from an optical modulation of the
counterpart of Aql X-1 \cite{che91,wel00}.  Despite several attempts,
the radio counterpart has been observed in only a few outbursts at a
level of ${\sim} 0.1-0.5$ mJy \cite{hje90, rup04, rup05c}.
In November 2009, radio observations provided evidence for the emission 
to arise from steady jets triggered at state transitions from the soft to 
hard state~\cite{mil10}. 

\subsection{SAX J1748.9--2021}

SAX J1748.9--2021 was discovered with the \textit{BeppoSAX}/WFC on
August 22, 1998 in the globular cluster NGC 6440
\cite{int99,ver00}. The distance of the source is well constrained
thanks to the knowledge of the globular cluster distance of
$8.5\pm0.4$ kpc \cite{ort94}. \textit{RXTE}/PCA and
\textit{BeppoSAX}/WFC monitoring revealed no pulsations, but
thermonuclear bursts were observed with no burst oscillations. The
position of the source was coincident with the position of MX 1746-20
detected by \textit{Uhuru} and \textit{OSO}-7 in 1971-1972. However,
given the large error box of these missions, it is not possible to
establish a secure association between the two sources.  SAX
J1748.9--2021 showed a typical average X-ray spectrum with a power-law
with photon index 1.5 and a soft blackbody component at $kT=0.9$ keV
\cite{int99}. However, other spectral models also gave a good fit to
the data \cite{int99}.

In August 2001, after 3 years of quiescence, SAX J1748.9--2021 started
a new outburst observed with the \textit{RXTE}/ASM.  \textit{Chandra}
observations allowed the establishment of a precise position in the
globular cluster and confirmed that the 1998 outburst was from the
same source. Thermonuclear bursts without burst oscillations were
again observed. The beginning of the third outburst was detected in
2005 March 7 during a routinely \textit{RXTE}/PCA Galactic Bulge Scan
Survey. The source was monitored until 2005 July 21 and several
pointed \textit{RXTE}/PCA observations were performed.  The first
pulsations reported for this source were seen in the 2005 outburst
\cite{gav06, gav07}. A coherent signal at $\simeq442$ Hz was detected
during a flux decay reminiscent of the tail of a superburst.  However,
neither the temporal profile nor the energetics of the tail were
consistent with a super-burst and this lead to the suggestion that
this was a new intermittent AMXP \cite{gav07}. An independent analysis
of \textit{RXTE}/PCA archival data from the 2001 and 2005 outburst
 revealed intermittent coherent X-ray pulsations at 442 Hz,
appearing and disappearing on timescales of few hundred
seconds~\cite{pat09, alt08}.  A further outburst was observed in
December 2009 with \textit{RXTE} and intermittent pulsations were
observed again~\cite{pat10e}.  Observations in quiescence carried out
in July 2000 and June 2003 revealed an X-ray spectrum with a thermal
component detected in both observations. The 2000 observation,
however, required also a power law component possibly because of
residual accretion on the NS~\cite{cac05}.

Thanks to \textit{ROSAT}/HRI archival observations and a 3.5 m New
Technology Telescope (NTT) observation on 1998 August 26--27, two
candidate optical counterparts, dubbed V1 and V2, were
identified~\cite{ver00}.   V2, with $B\simeq
22.7$ mag, has now been recognized as the true counterpart thanks to the
precise astrometric position obtained with the 2001 \textit{Chandra}
observations \cite{int01b}. VLA radio observations provided stringent
upper limits during the 2009 outburst and implied that the radio
emission is quenched at high X-ray luminosities~\cite{mil10b}. 

\subsection{NGC 6440 X--2}\label{sec:6440}

On 2009 June 28, the X-ray transient NGC 6440 X-2 was discovered
\cite{hei09b} in a \textit{Chandra} observation of the globular
cluster NGC 6440 (the same cluster that harbors the AMXP SAX
J1748.9--2021). Coherent pulsations at 206 Hz were detected in a
subsequent outburst in \textit{RXTE}/PCA observations on 2009 August
30~\cite{alt10}. The orbital period is 0.95 hr, making this the fifth
ultra-compact AMXP discovered. One of the most striking features of
this pulsar is the very low long-term average mass accretion rate
$\mdota<2\times10^{-12}\msun\rm\,yr^{-1}$, which could be determined
thanks to the well-defined distance to the globular cluster
\cite{hei10}. Peak flux is one of the faintest observed from the AMXPs
($L_{2-10\rm\,keV}<1.5\times 10^{36}\ergs$) and the outburst length is
extremely short, with an duration of $\approx$ 3--5 days.  It should
be noted that the \textit{RXTE}/PCA Galactic Bulge Scan flux limit
corresponds to a luminosity ${\sim} 10^{35}\ergs$~\cite{alt10} so that
the presence of a longer and very faint outburst cannot be ruled out,
although archival \textit{Chandra} observations did not reveal any
counterpart in quiescence with an upper limit on the X-ray luminosity
of $L_{X}\lesssim 10^{31}\ergs$. This raises the question of how many
faint sources of this kind might have been missed, particularly
important given that AMXPs seem to be associated with faint LMXBs, and
systems of this type may therefore contain undiscovered AMXPs.

The recurrence time of the outburst is also the shortest among AMXPs,
${\sim}$ one month. An \textit{RXTE}/PCA and \textit{Swift}/XRT
monitoring campaign revealed the occurrence of 10 outbursts between
2009 and 2011~\cite{hei10,pat13c}).  In November 2009, visibility
constraints and a new outburst of the other AMXP SAX J1748.9--2021
prevented further observations of NGC 6440 X-2.  A strong
(${\sim}50\%\rm\,rms$) 1 Hz modulation was observed in the light-curve
of at least 6 outbursts~\cite{pat13c}, which strongly resembled the 1
Hz modulation seen in SAX J1808.4--3658~\cite{vank00, pat09c}.
 

Despite the precise astrometric X-ray position of the source, the
optical counterpart of NGC 6440 X-2 has not yet been identified, with
$B>22$ and $V>21$ from archival \textit{Hubble Space Telescope}
(HST) imaging of the globular cluster when NGC 6440 X-2 was in
quiescence. During the 2009 August outburst, Gemini-South observations
did not reveal any counterpart with $g'>22$. Observations carried with
the CTIO 4-m telescope during the 2009 July outburst also did not
reveal any counterpart with $J>18.5$ and $K>17$ \cite{hei10}.

\subsection{IGR J17511--3057}

The X-ray binary IGR J17511--3057 was discovered during Galactic Bulge
Monitoring by the \textit{INTEGRAL} satellite in 2009 September
12~\cite{bal09}. Coherent pulsations at 245 Hz and an orbital
modulation of 3.5 hr were measured with follow up observations by the
\textit{RXTE}/PCA \cite{mar09b}. Given the X-ray mass function, the
minimum donor star mass is 0.13$\msun$.  The X-ray light-curve was
also extensively monitored by \textit{Swift}/XRT, \textit{XMM-Newton},
\textit{RXTE}/PCA and \textit{Chandra}. The light-curve showed a
typical fast rise with a slow decay lasting for about 20 days, before
entering a fast decay phase~\cite{pap10, alt10d}. Curiously, while the
source was in the fast decay phase, a very short and faint outburst of
the ultra-compact AMXP XTE J1751--305 appeared in the same field of
view (FoV) of \textit{RXTE}.  The mini-outburst of XTE J1751--305
lasted less than 3 days, and pulsations were clearly visible at 435 Hz
\cite{mar09}.  Thermonuclear bursts were observed in IGR J17511--3057
with \textit{Swift}~\cite{boz10}, and burst oscillations were detected
in \textit{RXTE} data \cite{wat09,alt10d}.  None of the bursts showed
PRE, so there are several different distance determinations in the
literature. The most constraining gives an upper limit on distance of
$\lesssim5$ kpc~\cite{alt10d}.

\textit{Swift} spectral analysis revealed a power law
($\Gamma{\sim}2$) plus a blackbody at 0.9 keV~\cite{boz10}. Broadband
spectral data from simultaneous \textit{RXTE}/PCA and
\textit{Swift}/XRT observations could be well fitted with a three
model component~\cite{ibr11}: a disk with low temperature
($kT{\sim}0.24$ keV), a hot black body (hot spot with $kT{\sim}1$ keV) and
a Comptonization component originating from the accretion shock
(electron temperature $T_e{\sim}30$ keV; $\tau_T{\sim}2$). The spectral
fitting also required a fluorescent iron line at 6.4 keV with Compton
reflection and provided an interstellar absorption column of
$N_{H}=0.88^{+0.21}_{-0.24}\times10^{22}\rm\,cm^{-2}$. Very similar
results were obtained using simultaneous \textit{XMM-Newton} and
\textit{RXTE}/PCA spectral data~\cite{pap10}.  Combined
\textit{RXTE}/PCA, \textit{Swift}/XRT and \textit{INTEGRAL} data
also gave consistent results using a thermal Comptonization
model~\cite{fal11}.  Twin kHz QPOs were observed at the beginning and
at the end of the outburst, and had $\Delta\nu\simeq120$ Hz, about half
the value of the spin frequency~\cite{kal11}.  A NIR counterpart with brightness $K_s=18.0\pm0.1$ was identified with
the 6.5 m Magellan Baade telescope on 2009 September 22
\cite{tor09}. A second observation on October 7, while the
X-ray flux was fading rapidly, showed no counterpart with 3$\sigma$
upper limits of $K_s>18.8$.  No radio counterpart has been found, with
upper limits of 0.10 mJy \cite{mil09}.

\subsection{Swift J1749.4--2807}

The first observation of Swift J1749.4--2807 was made on June 2, 2006
with the \textit{Swift}/BAT telescope, which recorded a high level
activity consistent with an unidentified source~\cite{sch06}.  This
was later identified as a ``burst-only'' accreting NS binary (i.e., a
NS that is bright enough to be detectable only during the occurrence
of thermonuclear explosions) thanks to a spectral analysis of the
\textit{Swift}/BAT data~\cite{wij09}. The analysis was consistent with
the observation of a thermonuclear burst for a source distance of
$6.7\pm1.3$ kpc. Further \textit{Swift}/XRT data analysis revealed an
X-ray counterpart of the burst and archival \textit{XMM-Newton} data
showed a faint point source coincident with the \textit{Swift}/XRT
position~\cite{hal06, wij09}.  Increased activity was reported in 2010
by \textit{INTEGRAL}, and was linked to the accretion powered emission
process~\cite{pav10, che10}.  Pulsations were discovered immediately
afterwards at 518 Hz in \textit{RXTE}/PCA follow-up
observations~\cite{alt10e, alt11}. A very strong second harmonic was
detected at 1036 Hz~\cite{boz10b} and orbital modulation was measured
at 8.82 hr, giving a minimum donor mass of about
0.6$\msun$\cite{bel10, str10}. Swift J1749.4--2807 was finally
identified as the first eclipsing AMXP~\cite{mar10} and the
inclination of the system was constrained to be
$i{\sim}74^{\circ}-78^{\circ}$~\cite{mar10,alt11}. The binary showed
three eclipses with a duration of $2172\pm13$ s, allowing the first
attempted detection of Shapiro delay effects in X-ray timing data for an object
outside the Solar System~\cite{mar10}.

To fit the 0.5--40 keV spectrum, an absorption column
density of $N_H=3.0\times10^{22}\rm\,cm^{-2}$ and a power-law with
spectral index $\Gamma\simeq1.7$ are required~\cite{fer11}.  The
absorption column is about 3 times larger than the expected Galactic
column density in the direction of the source.  The X-ray light-curve
showed an exponential decay with a \textit{Swift}/XRT non detection
after 11 days since its first 2010 observation.  Due to the crowded field (the source is on the Galactic plane), no single counterpart has been identified in
NIR counterpart searches~\cite{dav11}.   More than forty counterparts were identified in
the X-ray error circle with ESO's Very Large Telescope (VLT).

\subsection{IGR J17498--2921}

IGR J17498--2921 was discovered by \textit{INTEGRAL} on 2011 August
11~\cite{gib11}. Soon after, coherent X-ray pulsations were discovered
in \textit{RXTE}/PCA data at 401 Hz with an orbital period modulation
of 3.8 hr~\cite{pap11b}. Type I X-ray bursts were observed with
\textit{INTEGRAL}~\cite{fer11b} and \textit{RXTE}.  Burst oscillations
were detected~\cite{lin11c, cha12} and evidence of PRE placed the
source at a distance of 7.6 kpc~\cite{lin11c}. \textit{Swift}/XRT
observed the source returning to quiescence on 2011 September
19~\cite{lin11d}.  A candidate NIR counterpart of IGR J17498--2921 was
first detected in archival data taken with the 4-m VISTA telescope at
the Paranal observatory~\cite{gre11}. The optical counterpart was
detected on 2011 August 25 and 26 with the 2-m Faulkes Telescope
South~\cite{rus11}. The position detected was consistent with both the
\textit{Chandra} and the NIR counterparts.  However, further
observations suggested that the NIR/optical counterpart is a
foreground star aligned by chance with the X-ray
source~\cite{vandenberg11}. This counterpart was confirmed to be a
foreground source with observations taken at the 2.5-m Irenee du Pont
telescope. The analysis suggested that, given the distance and
extinction of IGR J17498--2921, the expected magnitudes are $R{\sim}28$
and $J{\sim}23.3$, both well above the limit reached by the optical/NIR
observations~\cite{tor11}.

\subsection{IGR J18245--2452}\label{sec:18245}

The most recent addition to the AMXP family was discovered by INTEGRAL
on 2013 March 28~\cite{eck13} during an X-ray outburst (20--100 keV
luminosity of $3\times10^{36}\ergs$) and is perhaps the most
spectacular of all the AMXPs. The source, at a distance of 5.5. kpc is
located in the core of the globular cluster M28~\cite{hei13}.
\textit{XMM-Newton} observations revealed an AMXP whose spin (254 Hz)
and orbital parameters were identical to those of a previously known
radio millisecond pulsar located in the same region (PSR J1824--2452I,
a.k.a. M28I)~\cite{pap13a}. This makes this source the first known
radio millisecond pulsar that has switched to an
AMXP state. The 0.5–-10 keV luminosity shows a very peculiar X-ray
flickering with flux variations of up to 2 orders of magnitude
happening within a few seconds\cite{pap13a,fer13}. The two flux states can
be described by two very different spectral models with power-law
index $\Gamma{\sim}1.7$ (high flux) and $\Gamma{\sim}0.7$ (low flux)
\cite{fer13}. This has been interpreted as evidence for the onset of a
propeller phase \textit{during} the outburst, rapidly
alternating with a normal accretion phase \cite{fer13}.
A similar flickering was also observed in quiescence\cite{lin13b} when
the source switches from a stable low luminosity state of ${\sim}10^{32}\ergs$
to a flickering state with luminosity  ${\sim}10^{33}-10^{34}\ergs$.

An iron $K_\alpha$ line was also observed in the \textit{XMM-Newton}
spectrum~\cite{pap13a}.  Thermonuclear X-ray bursts were observed by
\textit{Swift}/XRT and MAXI \cite{pap13c, lin13a,ser13} and burst
oscillations were later identified~\cite{pat13a}. Archival optical
observations obtained with \textit{HST} revealed a faint counterpart
on April 2009 and 2010. However, on August 2009 the \textit{HST}
detected a blue counterpart, brighter by ${\sim}2$ mag in several
filters (F390W$=20.37\pm0.06$, F606W$=19.51\pm0.04$, and
F656N$=17.26\pm0.04$)\cite{pal13,coh13}, with a strong $H_\alpha$
emission indicative of the presence of an accretion disk four years
before the 2013 outburst.  Radio observations performed with ATCA on
2013 April 5, show a bright radio continuum counterpart (0.62 and 0.75
mJy at 5.5 and 9 GHz, respectively)~\cite{pav13}.
The source was last detected in X-rays on 2013 May 1 and soon after it
turned back on in radio as a millisecond pulsar~\cite{pap13b}.

\section{Accretion Torques}\label{sec:4}

Once the donor star overflows its Roche lobe, gas flowing through
the inner Lagrangian point $L_1$ carries large specific angular
momentum and hence forms an accretion disk around the
NS.  The type of disk depends on the microphysical conditions governing the gas dynamics~\cite{gho78, gho79, gho79a, psa99}.  If gas pressure dominates, the disk will be geometrically thin and optically thick \cite{sha73,
  nov73, gho78} with material moving in Keplerian orbits with orbital frequency:
\begin{equation}
\nu_{K}=\frac{1}{2\pi}\sqrt{\frac{GM}{R^3}}\simeq 767 \mathrm{~Hz}\left(\frac{M}{1.4\msun}\right)^{1/2}\left(\frac{R}{20\rm\,km}\right)^{-3/2}
\end{equation}
At a distance of a few tens of kilometers from the NS, the gas orbits
several hundred cycles per second and flows almost undisturbed until
the magnetic field of the NS (that for AMXPs is of the order of
$10^8$~G) is strong enough to perturb its orbit. At the
magnetospheric-radius $r_m$, the kinetic energy of the free-falling
gas becomes comparable to the magnetic energy of the NS magnetosphere
:
\begin{eqnarray}\label{eq:rm}
r_m &=& \xi\,r_{A} = \xi \left(\frac{\mu^4}{2 G M \mdot^2}\right)^{1/7}\nonumber\\
&=& 35{\rm\,km}\,\xi\left(\frac{\mu}{10^{26}\rm\,G\,cm^{3}}\right)^{4/7}\times\left(\frac{\dot{M}}{10^{-10}\,M_{\odot}\,yr^{-1}}\right)^{-2/7}\left(\frac{M}{1.4\,M_{\odot}}\right)^{-1/7}
\end{eqnarray}
where $\dot{M}$ is the mass accretion rate at the inner disk
boundary, $\mu$ is the dipole magnetic moment of the NS and
 $r_A$ is the Alfv\'{e}n radius. This latter parameter is 
calculated assuming spherical accretion:
\begin{equation}
\frac{B^2\left(r_A\right)}{8\pi}\simeq\frac{1}{2}\rho\left(r_A\right)\upsilon^2\left(r_A\right).
\end{equation}
The parameters $\rho$ and $\upsilon$ are gas density and velocity,
respectively.  The term $\xi\approx0.3-1.0$ is a correction factor due
to the non-spherical geometry of the problem and is required because
the gas orbits in a disk rather than falling radially from every
direction. In the disk geometry, magnetic and fluid stresses balance when:
\begin{equation}
B_p B_{\phi} r^2= \dot{M} \frac{\partial(\upsilon_{\phi} r)}{\partial r}
\end{equation}
where $B_p$ and $B_{\phi}$ are the poloidal and toroidal components of
the NS magnetic filed, $r$ the radial coordinate measured from the NS
center and $\upsilon_{\phi}$ the azimuthal velocity of the plasma at  $r$.  If one assumes that the transition region $\Delta r$
connecting the unperturbed plasma flow far from the NS and the
magnetospheric flow is much smaller than $r_m$, then the above equation
takes the form:
\begin{equation}
B_p B_{\phi} r_m^2\Delta r= \dot{M}\upsilon_{\phi}r_m
\end{equation}
Once the gas reaches the transition region $\Delta r$, it stops
flowing in Keplerian orbits and starts to co-rotate with the
magnetosphere. The gas exchanges angular momentum with the
magnetosphere and changes the NS spin frequency. The NS is spun up if
its specific angular momentum is smaller than that of the accreting
gas, and otherwise spun down.  The spin-up/spin-down condition can be
thought of in terms of characteristic radii: if $r_m$ is smaller than
the radius where the Keplerian frequency equals the NS spin frequency
(the co-rotation radius $r_{co}$), the NS is spun up, otherwise it is
spun down. The co-rotation radius can be defined as:
\begin{equation}
r_{co} = 1683 \left(\frac{M}{1.4\,M_{\odot}}\right)^{1/3}\nu_{s}^{-2/3}\rm\,km.
\end{equation}
It is important to stress that this is an over-simplified picture of
the true physical conditions close in the inner disk.  In this simple
description, the accretion torque exerted on the NS for a Keplerian
disk truncated at $r_m$, with $r_m<r_{co}$, is:
\begin{equation}\label{eq:torque}
N = 2\pi I \nudot_s = \mdot\sqrt{G M r_m}
\end{equation}
where $I$ is the moment of inertia of the NS, $\nudot_s$ the NS spin
frequency derivative and $G$ the universal gravitational constant.  At
radii larger than the co-rotation radius, the magnetic field lines are
threaded into the accretion disk and dragged by the high conductivity
plasma so that an extra torque due to magnetic stresses has to be
expected~\cite{gho79, gho79a, wan87, rap04} in addition to the torques due to
the matter flow.  Spin down due to dipole emission is also always
present. A possible way to express the total torque acting on the pulsar
is~\cite{tau12}:
\begin{equation}\label{eq:torque2}
N = \left(\mdot\sqrt{G M r_m} + \frac{\mu^2}{9 r_m^3}\right)n(\omega) - \frac{\dot{E}_{dipole}}{2\pi\nu_s}
\end{equation}
where $\dot{E}_{dipole}$ is the energy loss due to dipole radiation
and $n(\omega)\approx\pm\,1$ is a dimensionless function that depends
on the fastness parameter $\omega=\left(r_m/r_{co}\right)^{3/2}$:
\begin{equation}
n\left(\omega\right) = \rm\,tanh\left(\frac{1-\omega}{\Delta\,r}\right)
\end{equation}
This term takes account of the gradual transition from the spin-down to
spin-up zone in the accretion disk~\cite{rap04, tau12}.  The two extra
terms in Eq.(\ref{eq:torque2}) have a minor effect during most of the
outburst, when the mass accretion has the largest weight in
determining the net torque. The expression in Eq.(\ref{eq:torque}) is
therefore a good approximation most of the time.

\begin{figure}[t!]
\sidecaption
\rotatebox{0}{\includegraphics[scale=.42]{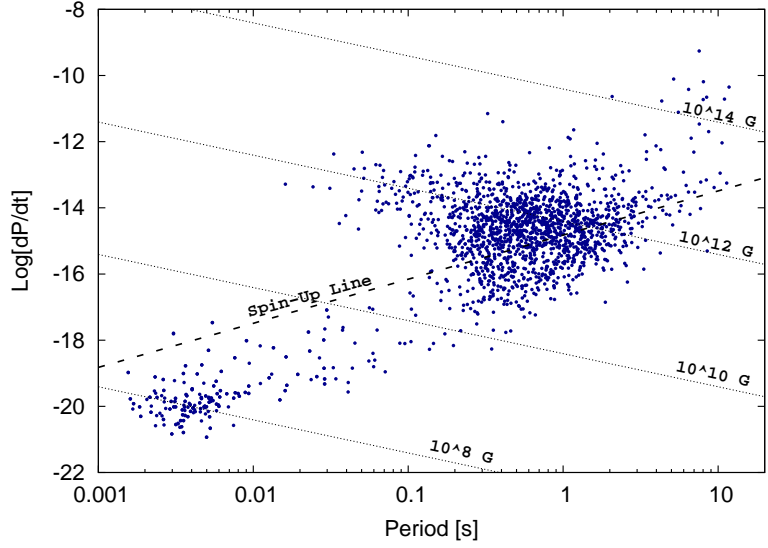}}
\caption{$P-\dot{P}$ diagram of radio pulsars with a spin up line
  (dashed black line) above which radio pulsars should not be found if
  they are ``recycled'' via accretion. The black dotted lines
  identify magnetic fields of different strengths.}
\label{fig:ppdot}       
\end{figure}

As the AMXP is spun up, $r_{co}$ moves towards and eventually reaches
$r_m$. When this happens, the pulsar is said to have reached the
``equilibrium spin period'' $P_\mathrm{eq}$:
\begin{equation}\label{eq:eq}
P_{\rm\,eq} \simeq 2.7 \left(\frac{B}{10^8 {\rm\,G}}\right)^{6/7}\left(\frac{M}{1.4\,M_{\odot}}\right)^{-5/7}\left(\frac{\dot{M}}{10^{-10}\msun\rm\,yr^{-1}}\right)^{-3/7}\left(\frac{R}{10 {\rm\,km}}\right)^{18/7} {\rm\,ms}
\end{equation}
Substituting into Eq.(\ref{eq:eq}) the surface magnetic field
(at the poles) derived from dipole spin down~\cite{lor04}: 
\begin{equation}
B=\sqrt{\frac{6 c^3 I P\dot{P}}{4 \pi^2 R^6}}\frac{1}{\rm\,sin\it\alpha} \simeq 6.4\times10^{19} {\rm\,G}\sqrt{P\dot{P}}\left(\frac{M}{1.4\,M_{\odot}}\right)^{3/2}\frac{1}{\rm\,sin\it\alpha}
\end{equation}
(where we have assumed $I=10^{45}\rm\,g\,cm^{2}$, $R=10$~km and
$\alpha$ is the misalignment angle between spin and magnetic axes),
one obtains a relation between $P_{\rm\,eq}$ and $\dot{P}$.  When the
accretion rate reaches the maximum Eddington rate, this relation
defines a ``spin-up line'' in the $P-\dot{P}$ diagram of radio pulsars
(see Figure~\ref{fig:ppdot}), above which millisecond radio pulsars
should not be found. Indeed, if millisecond pulsars are created in
LMXBs via accretion torques, then the maximum possible torque is set
by the Eddington limit\footnote{Note, however, that the spin-up line
  depends on several parameters which are difficult to constrain like
  the angle $\alpha$.  Its position in the $P-\dot{P}$ diagram is
  therefore subject to uncertainties (see~\cite{tau12b} for a
  discussion).}  It is important to stress that the numerical solution
of the force-free relativistic MHD equations lead to a similar (but
slighly different) result for the dipolar magnetic field (again, at
the poles) \cite{spi06}:
\begin{equation}\label{eq:B}
B=\sqrt{\frac{c^3 I P\dot{P}}{\pi^2 R^6}}\frac{1}{\left(1+\rm\,sin^{2}\it\alpha\right)^{1/2}} \simeq 5.2\times10^{19} {\rm\,G}\sqrt{P\dot{P}}\left(\frac{M}{1.4\,M_{\odot}}\right)^{3/2}\frac{1}{\left(1+\rm\,sin^{2}\it\alpha\right)^{1/2}}
\end{equation}

To detect the effect of accretion torques in AMXPs one simply needs a
measurement of $\nudot_s$, which in principle is straightforward when
using coherent timing techniques. Unfortunately, however, determining the
location of the magnetospheric radius is a non-trivial problem.
\begin{svgraybox}
The main reason is that $\Delta r/r_m$ is assumed to be much smaller
than unity, which is a good approximation only for slow accreting
pulsars with high magnetic fields. This might not be the case for
AMXPs with rapid rotation and weak magnetic fields (see for example
Eq. 5 and the underlying assumptions).
\end{svgraybox}
Furthermore we have considered so far only gas-pressure dominated
accretion disks, considered to be a good assumption at
the low average accretion rates of AMXPs. However, when the
outbursts are close to their peak luminosities, radiation pressure
may play a role in the exchange of angular momentum between the gas
and the NS. At these accretion rates accreting plasma and
magnetic field lines may couple and modify the amount of enhanced
angular momentum~\cite{and05, pat12a}, thus invalidating the use of
Eq.(\ref{eq:torque}-\ref{eq:torque2}).

Noting these caveats, theoretical expectations for the
average spin frequency derivative of an AMXP can be calculated with
the simplest accretion model (Eq.~\ref{eq:torque}) once one has a
reasonable estimate of the mean mass accretion rate $\mdot$ \textit{during
an outburst}. As discussed, such a measure is difficult to
obtain, but several estimates of $\mdot$ are present in the
literature which can be taken as a first step to compare the measured
$\dot{\nu}_s$ with the expected values (see for example \cite{hei09}
and \cite{pat12a}). The explicit expression for the expected
$\nudot_s^{exp}$ can be obtained by substituting Eq.(\ref{eq:rm}) into
Eq.(\ref{eq:torque}). 
The substitution gives:
\begin{equation}
\nudot_s^{\rm\,exp} = 2.3\times10^{-14} \xi^{1/2} \mdot_{-10}^{6/7}M_{1.4}^{3/7}B_{8}^{2/7}R_{10}^{6/7} \rm\,Hz\,s^{-1}
\end{equation}
where we have normalized all variables with their typical values as in Eq.(\ref{eq:eq}). 

As one can easily verify, the calculated values of $\nudot_s^{exp}$
are all of the order of $10^{-13}\hzs$ when neglecting the fact that
at low mass accretion rates, when $r_m>r_{co}$ the ``propeller
regime'' can set in, with the spin-down terms dominating in
Eq.(\ref{eq:torque2}).  In this phase, the specific angular momentum
of the accreting plasma is insufficient for spin up to occur, and the
centrifugal barrier removes angular momentum from the AMXP, slowing it
down. It was originally suggested that centrifugal inhibition by the
rotating magnetosphere would expel gas from the system and shut down
the accretion process~\cite{ill75}. In fact for this to happen,
$r_{m}$ must exceed $r_{co}$ by a margin of at least 1.3 so that
matter can be accelerated to the escape velocity and can be flung out
of the disk~\cite{spr93,rap04,dan10}.  Accretion can in fact still
take place, and two propeller regimes have been observed in 3D-MHD
simulations of accreting pulsars \cite{rom05, ust06}: a ``weak
propeller'' with no outflows and a ``strong propeller'' with expulsion
of material.  In both cases channeled accretion is still ongoing so
that a spinning down AMXP could be in principle observed via coherent
timing measurements in either propeller regime. The propeller is
therefore expected to affect $\nudot_s$, but determining its onset is
a very difficult task.  Indeed, AMXP observations do not help much to
constrain the theory in this case, since only weak evidence for
possible propeller phases exists (\cite{pat09c, pat13c, har11,
  fer13}). Coherent timing does not provide further insights, since
the propeller is expected to start when the mass accretion (and thus
the source luminosity) drops substantially, with a consequent decrease
of the S/N of the pulsations. Furthermore the propeller may not last
sufficiently long to allow measurement of the spin frequency
derivative. 

\subsection{Coherent Timing Technique}\label{4.1}

Coherent timing analysis of the phase evolution of AMXPs can be
performed after converting photon times of arrival from the spacecraft
non-inertial reference frame to the Solar System barycenter (an
approximate inertial reference frame) and correcting for general and
special relativistic perturbations due to planets and minor bodies in
the Solar System that affect photon propagation.  Before discussing
observational tests of accretion theory it is useful to clarify the
observables that play a role in coherent timing studies. AMXPs show
pulsations that are too weak to detect as single pulses: to
reconstruct the signal it is necessary to fold the data in segments of
several hundred seconds to obtain a pulse profile that is the average
of several hundred thousand NS cycles. It is then possible to measure
the fractional amplitude of the pulsations, and the pulse phase. Pulse
profiles of AMXPs are generally highly sinusoidal, with little or no
harmonic content beyond the fundamental frequency.  In this case the
pulse phases can be measured by choosing a fiducial point (e.g. the
pulse peak) and tracking the variation of the phase at this point over
time. Sometimes, however, strong harmonic content is observed, with
the pulse profile shape varying during an outburst.  This means that
unlike in radio pulsar timing, where average profiles are often
very stable, there is no stable fiducial point on which to base timing
analysis.  To avoid this problem it has become standard to decompose
the pulse profile into its harmonic components and measure pulse
amplitude $b_k$ and phase $\phi_k$ of the $k$-th harmonic ($k=1$ for
the fundamental, $k=2$ for the second harmonic and so on) via the
expression:
\begin{equation}
x_j = b_0 + \sum_k b_k{\rm\,cos\it{\left[2\pi\left(\frac{\rm\,k(j-0.5)}{N}-\it{\phi_k}\right)\right]}}
\end{equation}
where $b_0$ is the unpulsed component, $x_j$ is the number of counts
detected in the $j$-th bin of the pulse profile, with $j=1,2,...,N$.
This way the fiducial point of each harmonic is well defined since
all harmonic components are pure sinusoids.  The fractional amplitude
can be measured by adding in quadrature the fractional amplitudes of
each harmonic:
\begin{equation}\label{fa}
R = \left(\sum_k R_k^2\right)^{1/2} = \sum_k \frac{N b_k}{N_{ph} - B}
\end{equation}
where $R_k$ is the fractional amplitude of the $k$-th harmonic,
$N_{ph}=\sum_j x_j$ the total number of photons in a profile and $B$
the total number of background photons. Note that this definition
gives a fractional amplitude larger by a factor $\sqrt{2}$ than the
often reported rms fractional amplitude. We use this definition of
fractional amplitude since it has an immediate physical meaning as the
pulsed flux fraction. To avoid confusion, we always refer to the
fractional amplitude as ``sinusoidal fractional amplitude'' when we
use the definition given in Eq.(\ref{fa}) or otherwise to ``rms
fractional amplitude''. Note that in the AMXP literature ``rms
fractional amplitude'', ``sinusoidal fractional amplitude'' and
``peak-to-peak fractional amplitude'' are also used.  This latter is
calculated by measuring the flux at the peak and at the minimum of the
pulse and dividing by the average flux.

Pulse phase can be measured as a function of time and each
harmonic analyzed separately. We drop the subscript $k$
from the phase symbols, since the following equations are
valid for each harmonic. The pulse phase series can be
decomposed into several terms encoding different physical effects:
\begin{equation}
\phi\left(t\right) = \phi_{L} + \phi_{Q} + \phi_{orb} + \phi_A + \phi_M + \phi_N
\end{equation}
\begin{itemize}
\item $\phi_{L} = \phi_0 + \nu t$ is the linear term due to a \textit{constant} spin of the NS ($\phi_0$ is an initial reference phase)
\item $\phi_Q = \frac{1}{2}\nudot t^2$ is the quadratic variation due to \textit{constant} spin up or spin down
\item $\phi_{orb}$ contains the effect of the
orbital motion of the NS around the companion and, as a first order approximation, can be calculated by 
measuring the delays in the time of arrival of photons:
\begin{equation}
t_{em} = t_{arr}-\frac{A_1}{c}\rm\,sin\it{i}\rm\,sin\it\left[\frac{2\pi}{P_b}\left(t_{arr}-T_{asc}\right)\right]
\end{equation}
where $A_{1}$ is the semi-major axis of the NS orbit, $c$ the speed of
light, $i$ the inclination of the orbit with respect to the observer,
$T_{asc}$ the time of passage through the ascending node, $P_b$ the
orbital period and $t_{em}$ and $t_{arr}$ the photon emission and
arrival time \cite{pap05}. In this expression we have assumed that the
orbit is perfectly circular, a good first order approximation in all
AMXPs. More sophisticated models for nearly circular orbits can be
found for example in \cite{lan01}.
\item $\phi_A$ gives phase variations related to uncertainty in
  astrometric source position, which introduces a spurious frequency
  and frequency derivative offset. These offsets can be expressed as
  (\cite{har08}):
\begin{eqnarray}
\Delta\nu &=&
\nu_0\epsilon\left(a_{\oplus}\rm\,cos\it{\beta}/c\right)\left(2\pi/P_{\oplus}\right)\rm\,cos\it{\tau}\\ \Delta\nudot
&=&
-\nu_0\epsilon\left(a_{\oplus}\rm\,cos\it{\beta}/c\right)\left(2\pi/P_{\oplus}\right)^2\rm\,sin\it\tau
\end{eqnarray}
Here $\nu_0$ is the true pulse frequency, $\epsilon$ the
position error parallel to the plane of the ecliptic, $\beta$ the
ecliptic latitude of the AMXP, $a_{\oplus}$ and $P_{\oplus}$ the
Earth semi-major axis and orbital period and $\tau=2\pi\,t/P_{\oplus}$ the orbital phase of the Earth.  Phase zero is defined as the point where the Earth is closest to the
AMXP and for order of magnitude estimates can be assumed such that
${\rm\,cos}\tau ={\rm\,sin}\tau = 1$.
\item $\phi_M$ refers to unavoidable phase wandering due to
measurement errors, and is normally distributed with an amplitude
predictable by propagating the Poisson uncertainties due to counting
statistics.  
\item $\phi_N$ is a subtle term covering
  residual phase variations that do not fall into any of the previous categories. These are usually called ``X-ray timing noise'',
  by analogy with the timing noise often observed in radio pulsars.
  Note that if the spin-up (or spin-down) process is not constant in
  time we \textbf{do expect} terms higher than the quadratic
  ($\phi_Q$) and these are considered, in our definition, as part of
  $\phi_N$ even if they are true variations of the NS rotation. This
  should be expected if, for example, the accretion torques exhibit
  stochastic variations.
\end{itemize}

When performing coherent timing after having removed the $\phi_{orb}$
term and assuming that $\phi_A$ is negligible, the observer measures a
pulse frequency and its time derivatives.  These can be determined
using a Taylor expansion:
\begin{equation}
\phi\left(t\right) = \phi_0 + \frac{\partial\phi}{\partial t}(t-t_0) + \frac{\partial^2\phi}{\partial\,t^2}\frac{(t-t_0)^2}{2!} + ...
\end{equation}
The parameters $\nu=\frac{\partial\phi}{\partial t}$, $\nudot=\frac{\partial^2\phi}{\partial\,t^2}$, etc., can be determined by fitting the pulse
phases with standard $\chi^2$ minimization techniques. 
\begin{svgraybox}
The \textit{observable} quantities here are \textit{not} the spin
frequency $\nu_s$ and derivatives, but the pulse frequency $\nu$ and
derivatives which are encoded as a combination of $\phi_{L}$, $\phi_Q$
and $\phi_N$. The pulse frequency is the frequency of the pulsations
detected by the distant observer, whereas the spin frequency is the
rotational rate of the NS as measured by the distant observer. The
assumption that pulse and spin frequency (and derivatives) are
identical may not always be true. For pulse and spin frequencies to be
identical, $\phi_N$ must have no linear component. Similarly for the
pulse and spin frequency derivative: only if $\phi_N$ has no quadratic
component will the two be the same.
\end{svgraybox}

\subsection{Observations: Accretion Torques in AMXPs}

With the exception of the intermittent pulsar Aql X-1, all of the
AMXPs have shown pulsations of sufficient quality and with a
sufficiently long baseline to constrain the pulse frequency
derivatives. As a rule of thumb, the condition that must be met to
detect a pulse frequency derivative $\nudot$ in a data segment of
length $\Delta t$ is that $\sigma_{rms}<\nudot\left(\Delta
t\right)^2$, where $\sigma_{rms}$ is the root-mean-square error on
pulse phases. Since all AMXPs (bar Aql X-1) have $\sigma_{rms}{\sim}
0.01$ cycles, and baselines of several days, we are sensitive to
$\nudot {\sim} 10^{-15}-10^{-13}\hzs$.  These values overlap the range
of expected $\nudot$ given by Eq.(\ref{eq:torque}) for typical $\mdot$
and weak dipolar $B$ fields ${\sim} 10^8-10^9$ G.

Several papers have reported pulse frequency derivatives in AMXPs. The
first measure was made for the source IGR J00291+5934~\cite{fal05b}
with a reported $\nudot=8.4\times10^{-13}\hzs$ during its 2004
outburst.  Table~\ref{tab:nudotp} summarizes the measurements for all
AMXPs.  However none of these values, including the measurement made
for IGR J00291+5934, take into account the presence of X-ray timing
noise in the pulse phases.  One must therefore bear in mind that what
is reported is the combined effect of $\phi_Q$ and $\phi_N$, as
explained in Section~\ref{4.1}. Nevertheless timing noise has
different strength in different sources, so that pulse and spin
frequency derivatives may in some cases not be so dissimilar. We have
marked each AMXP in Table~\ref{tab:nudotp} with a ``w'' for weak
timing noise and ``s'' for strong.  This distinction is somehow
arbitrary, but is useful to understand the probability of the pulse
frequency derivatives differing from the true spin frequency
derivatives.

\begin{table}
\caption{Pulse Frequency and Spin Frequency Derivatives}

\scriptsize
\begin{center}
\rowcolors{3}{gray!10}{}
\begin{tabular}{llclll}
\hline
\hline
Source & Pulse freq. deriv. $\nudot$& Timing noise & PFD            &    Spin freq. deriv  $\nudot_s$& SFD         \\
              & (PFD) $[\rm\,Hz\,s^{-1}]$                   &  strength         & Reference &   (SFD) [$\rm\,Hz\,s^{-1}$]                 & reference \\
\hline
SAX J1808.4--3658 & $4.4\times10^{-13}$; $-7.6\times10^{-14}$ & s  & \cite{bur06} & $<|2.5|\times10^{-14}$ & \cite{har08, har09}\\
XTE J1807--294 & $2.5\times10^{-14}$ & s & \cite{rig08} & $<|4|\times10^{-14}$& \cite{pat09b} \\
\smallskip
IGR J00291+5934 & $[5; 11]\times10^{-13}$ & w & \cite{fal05b,bur07,pap11} & $+5.5\times10^{-13}$ &\cite{fal05b, bur07, pat10b,
  har11, pap11} \\
XTE J1814--338 & $-6.7\times10^{-14}$ & s & \cite{pap07} & $<|1.5|\times10^{-14}$ & \cite{has11} \\
XTE J0929--314 & $-9.2(4)\times10^{-14}$ & s & \cite{gal02} & &  \\
\smallskip
XTE J1751--305 & $5.6\times10^{-13}$& w & \cite{pap08} & & \\
IGR J17511-3057 & $1.6\times10^{-13}$ & s & \cite{rig11c}  & & \\
IGR J17498--2921 & $-6.3\times10^{-14}$ & w & \cite{pap11b} & & \\
\smallskip
HETE J1900.1--2455 & & s & & $[2.3;0.4]\times10^{-13}$ & \cite{pat12c}  \\
Swift J1756.9-2508 & & w & &                                                       $<|3|\times10^{-13}$ & \cite{pat10c}\\ 
Swift J1749.4-2807 & & s & & $<|1.2|\times10^{-12}$ & \cite{mar10}\\
IGR J18245--2452 & & ? & &  $<|1.3|\times10^{-12}$ & \cite{pap13a}\\
\end{tabular}\label{tab:nudotp}
\end{center}
The AMXPs Aql X--1, SAX J1748.9--2021 and NGC 6440 X--2 do not appear
in the table above as the observting baseline is too short to provide
meaningful upper limits.
\end{table}

To verify whether the pulse frequency derivative is a robust indicator
of the spin frequency derivative $\nudot_s$, one has to investigate
the strength of $\phi_N$ on the timescales over which $\nudot$ is
measured. To do this, one can use Monte Carlo (MC) simulations to
estimate the true uncertainty on the pulse frequency derivative
\cite{har08}. The MC method works as follows: one fits a second-order
polynomial to measure $\nu$ and $\nudot$ with standard $\chi^2$
minimization techniques after having removed all other effects (such
as orbital variations) and obtains an estimate of $\nu\pm\sigma_{\nu}$
and $\nudot\pm\sigma_{\nudot}$.  If in the phase residuals there are
still substantial variations of the pulse phase, in excess of that
expected from measurement errors alone (i.e., $\phi_M$) - which can be
verified by checking whether $\chi^2$ gives a statistically
unacceptable fit - then the statistical errors $\sigma_{\nu}$ and
$\sigma_{\nudot}$ are not good representations of the true
uncertainties of the spin parameters.  Instead one can take the phase
residuals, calculate a power spectrum and simulate several thousand
time series with nearly identical noise content as the original phase
residual time series~\cite{har08}.  At this point $\nu$ and $\nudot$
can be measured for each simulated time series, and a distribution of
parameters constructed. The standard deviation of the distribution of
$\nu$ and $\nudot$ provides a good representation of the true
uncertainties on the spin parameters.  In this way one can immediately
check whether the noise content $\phi_N$ affects the measured value of
$\nu$ and $\nudot$. Applying this technique has revealed discrepancies
between pulse and spin parameters in several AMXPs
(Table~\ref{tab:nudotp}).  The most striking finding has been the
non-detection of spin frequency derivatives in several AMXPs, with
upper limits below the expected $\nudot_s$.

The use of MC simulations is not the final word on this problem, since
the origin of timing noise remains unexplained. It also has its
limitations: if the lowest Fourier frequency component of timing noise
is comparable to the length of the data segment over which $\nu$ and
$\nudot$ are measured, MC simulations cannot distinguish pulse and
spin parameters. It has been noted that the phase residuals obtained
after removing $\nu$ and $\nudot$ from the pulse phases of two AMXPs
(XTE J1814-338 and XTE 1807-294) are anti-correlated with variations
in X-ray flux \cite{pap07, rig08}. The anti-correlation improves
substantially if one fits a simple $\nu=const$ model, suggesting that
timing noise, which is related to the X-ray flux variations, is almost
entirely responsible for the measured $\nudot$
\cite{pat09f}. Correlations (or anti-correlations) between pulse phase
residuals (with respect to a $\nu=const$ model) and X-ray flux have
now been found in at least six AMXPs where such studies have been
carried out~\cite{pat09f}. In some cases the correlation was striking
(e.g., in XTE J1814-338~\cite{has11}) leaving little doubt that the
pulse frequency derivative $\nudot$ is \textbf{not} the spin frequency
derivative $\nudot_s$.  This discovery also suggests that \textbf{it
  is the pulse phase $\phi$ and not its second derivative ($\nudot$)
  that correlates with the flux} (and thus the mass accretion rate),
as would instead be expected from Eq.(\ref{eq:torque}).  One way to test these
findings is to measure ``instantaneous'' short timescale spin
frequencies and check whether these scale with the bolometric flux as
$\nudot_s\propto F_{bol}^{\gamma}$, where $\gamma$ is a scale factor
that depends on accretion disk structure.  The bolometric flux is,
however, almost never available, since observations usually cover only
a narrow (high) energy band. If we assume that the X-ray flux is
$F_{X}\propto F_{bol}$ then we expect to see $\nudot_s\propto
F_{X}^{\gamma}$. However this assumes that the mass accretion rate
$\mdot\propto F_{X}$, which has been shown to be untrue in some
LMXBs~\cite{van01}. There is one additional complication: AMXPs have
outbursts that are almost always too short to split data segments long
enough to yield more than one short-term ``instantaneous'' spin
frequency derivative.

If one keeps these caveats in mind, then it is possible to test the
relation between $\nudot_s$ and $F_{X}$ for some AMXPs, such as XTE
J1807--294 and XTE J1814--338 which have shown long outbursts (${\sim}100$
and ${\sim}40$ days, respectively) and high S/N ratio for the
pulsations.  In XTE J1807--294, the ``instantaneous'' $\nudot$ changes
sign several times during the outburst decay and all ``instantaneous''
spin frequency derivatives $\nudot_s$ are insignificant and consistent
with being part of the underlying timing noise process \cite{pat09b}.
In XTE J1814--338, the instantaneous $\nudot$ are too large (up to
$10^{-11}\hzs$) to be physically meaningful, requiring accretion rates
well above the Eddington limit \cite{wat08}. This is at odds with the
fact that AMXPs are faint and rarely reach accretion rates above
$10\%$ Eddington, suggesting that $\nudot$ is different to $\nudot_s$
and is completely dominated by timing noise. This does not prove that
$\nudot_s$ does not scale with flux, but only that measurements of
``instantaneous'' $\nudot_s$ are contaminated by timing noise, which
prevents testing of the relation between $\nudot_s$ and $F_{X}$.

Theoretical expectations for $\nudot_s$ mostly exceed observed values
measured by taking into account the contamination of timing noise
(Table~\ref{tab:nudotp}). This means that most AMXPs do not behave in
accordance with the predictions of accretion torque theory
(Eq.\ref{eq:torque2}). In addition, once gas attaches to the magnetic
field lines of the AMXP, an exchange of angular momentum is inevitable
unless $r_{m} = r_{co}$.  But this cannot be the case throughout an
outburst: $r_{m}\propto \dot{M}^{-2/7}$ and the mass accretion rate
varies, so $r_m$ must at some point differ from $r_{co}$. Larger
torques, which are expected close to the outburst peak, are clearly
ruled out in some AMXPs~\cite{har08, pat10, has11}.  The angular
momentum of the accreting plasma must however be transferred somewhere
so that the mismatch between expected and observed $\nudot_s$ appears
problematic.

\subsubsection{The Origin of X-ray Timing Noise}\label{noise}

\begin{figure*}[b!]
   \includegraphics[scale=0.18]{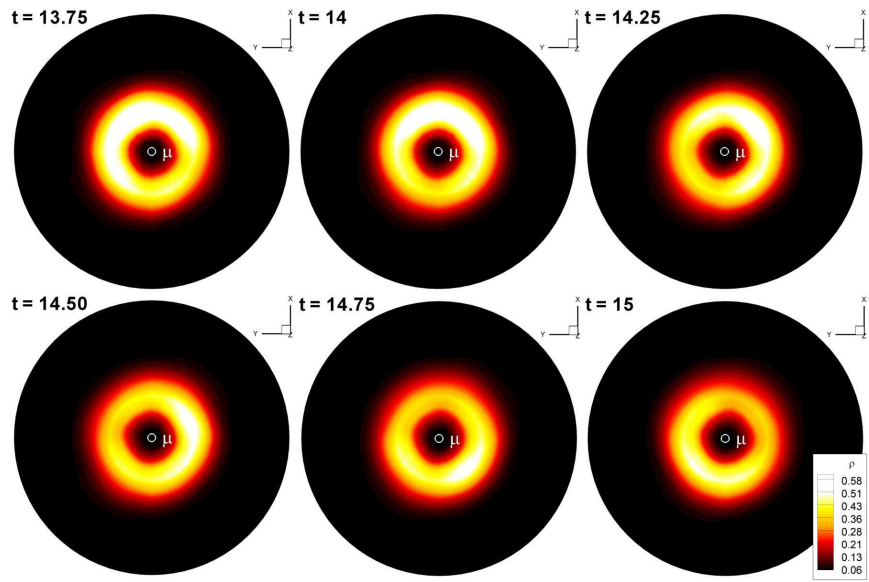}
\caption{Evolution of the position of the hot spot during accretion
  ($\nu_s=4.1$ms). $\mu$ represents the magnetic dipole axis. Note how
  the densest portion of the circular hot spot (encoded by the $\rho$
  scale, bottom right) shifts during time even if the observer is in
  the rotating reference frame of the NS (Figure from \cite{bac10}.)}
\label{fig:bachetti}
\end{figure*}

The phase wandering of pulsations in coherent timing analysis requires
careful consideration if we are to understand how a NS responds to
accretion. Theoretically the phenomenon is easy to explain if one
assumes that the accretion hot spot is not completely anchored to one
location on the surface.  One possibility is that the magnetic and
rotational axes of the NS are almost aligned and the hot spot wanders
around the magnetic axis by small displacements, generating large
variations in phase and amplitude ~\cite{lam09} . In this case pulsed
fractions and times of arrivals should be anti-correlated.  A moving
hot spot has been observed in 3D-MHD simulations of magnetized
accreting NSs \cite{rom04,bac10}.  In particular, it was shown that
the usual assumption of a fixed hot spot is valid only for large
misalignment angles between the magnetic and spin axes
(Figure~\ref{fig:bachetti}, \cite{bac10}), in good agreement with the
magnetic and rotational axis alignment model \cite{rud91,lam09}. This
strengthens the idea that most X-ray timing noise may be related to a
moving hot spot configuration.  Alternatively, the phase wandering
might be related to complexity in the structure of the pulsar's
magnetic field, with different multipole components dominating the
accretion process at different accretion rates~\cite{lon12}.

Timing noise may also be due to fluctuating accretion torques
\cite{lam78, lam78b}.  This has been studied for strongly magnetized
accreting pulsars like Vela X--1 \cite{dee89}, and it is natural to
expect something similar in AMXPs. However as discussed, this model
requires unphysical large accretion rate fluctuations to explain all
timing noise or the large jumps in phase observed for example in XTE
J1814--338 (and probably also SAX J1808.4--3658 and XTE J1807--294,
\cite{pap07, rig08, pat09f, wat08, har08}).  Nonetheless some X-ray
timing noise, or even its entirety for AMXPs with weak timing noise
content, could in principle be related to accretion torque
fluctuations.

\section{Pulse Profiles}\label{sec:5}

\begin{figure}[t!]
\sidecaption
\rotatebox{0}{\includegraphics[scale=.3]{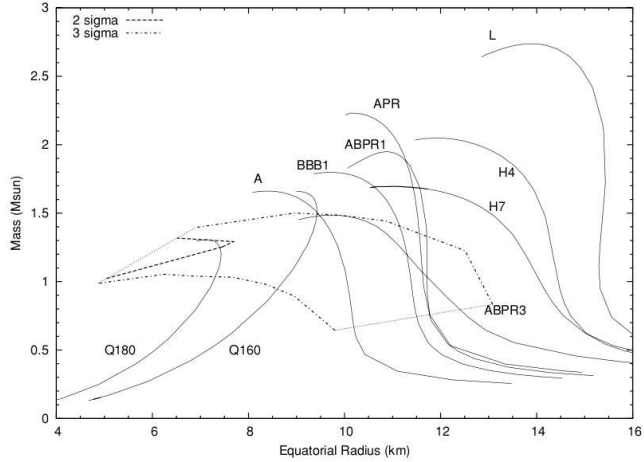}}
\caption{Mass-Radius relation for several different EoS of ultra-dense matter,
and constraints on the M-R relation of the AMXP SAX J1808.4--3658. These result from modeling pulse profiles observed in the 1998 and 2002 outbursts (Figure from~\cite{mor11}).}
\label{fig:eos-pulse}       
\end{figure}

AMXP pulse profiles form in regions of strong gravity and encode
information about the physical properties of the emitting regions and
the compactness of the NS \cite{pou03, pou08, lea08, lea11, mor11}.
Light bending, aberration and relativistic Doppler shifts all affect
pulse profile shape~\cite{cad07, mun02b, pou06, pou06b} and can be
measured using coherent timing analysis. This can be used to
constrain the EoS of ultra-dense matter (Figure~\ref{fig:eos-pulse})
although large uncertainties still exist: in part due to
model-dependencies, but more importantly because of the low S/N of the
relativistic features.

Pulse profile shapes can also constrain the geometry of the NS
magnetic field \cite{yah80, bul95, kra96} since multipoles generate
different accretion columns and hot spots on the NS surface (see,
however, \cite{ann10}, for a recent discussion of the problem).  The
information contained in pulse profiles can be extracted by studying
pulse amplitudes, the harmonic content of the signal and the time of
arrival of the pulsations. Double peaked pulse profiles, for example,
may be observed when two antipodal hot-spots exist on the NS surface
\cite{mun02b, lam09, pou06b}; a high harmonic content is suggestive of
a complex field geometry \cite{lon08} and the energy dependence of the
pulse amplitudes and the time lags of the pulsations provide a way to
explore the angular pattern of the radiation emitted from the NS
surface \cite{pou03, gie05, pou06b}. The temporal evolution of the
pulse shapes also reveals details of the complex disk/magnetospheric
interaction responsible for magnetic channeling of the accreting
plasma \cite{kul05, ibr09}.

\subsection{Pulse Fractional Amplitudes and Phase Lags}\label{sec:amp}

The pulse profiles of AMXPs are sinusoidal, with only the fundamental
frequency detected in most cases (including IGR J00291+5934, XTE
J0929--314, XTE J1751--305, SAX J1748.9--2021 and IGR
J17498--2921). In some AMXPs there is a first overtone (e.g. XTE
J1814--338, IGR J18245--2452 and NGC 6440 X-2), and higher overtones
are seen only sporadically in just three AMXPs (SAX J1808.4--3648,
Swift J1749.4--2807 and XTE J1807--294).  Pulse fractional amplitudes
reach values of a few percent, with the highest sinusoidal amplitude
measured being 30-40\% (Swift J1749.4--2807 and XTE J1807--294) and
the lowest being 0.3\% (HETE J1900.1--2455). With the exception of
Swift J1749.4--2807, overtones do not generally contribute to the
amplitude of the pulsations by more than about ${\sim}5\%$.

All AMXPs show an energy dependence of the pulse fractional amplitudes.
In some sources the fractional amplitude increases steeply with energy,
e.g. Aql X-1~\cite{cas08} Swift J1756.9--2508~\cite{pat10} and SAX
J1748.9--2021~\cite{pat09}. In this latter source, the
fractional amplitude increases by a factor ${\sim} 10$ between 2 and
${\sim}20$~keV. Some AMXPs have fractional amplitudes that drop with
energy (SAX J1808.4--3658~\cite{cui98}, XTE J1751--305~\cite{gie05},
XTE J0929--314~\cite{gal07}, HETE J1900.1--2455~\cite{gal07}, IGR
J17511--3057~\cite{ibr09, fal11}).  Other AMXPs (IGR J00291+5934
and XTE J1807--294) are more complex, with the
amplitude rising and falling in different energy
bands~\cite{fal05b,pat10}. The energy dependence of the pulsed
fractions provides insights into the pulse formation process, since a
simple hot spot emitting blackbody radiation with a temperature
contrast with respect to the NS surface would produce
pulsations which increase in amplitude at higher energies in the
observer reference frame~\cite{mun02b}.

Another interesting finding has been the correlation or
anti-correlation between the fractional amplitude of the fundamental
(where most of the pulse power is observed) and the time of arrival of
the pulsations of some AMXPs. The most prominent example is XTE
J1807--294 where high/low fractional amplitude pulses arrive
systematically earlier/later than predicted by the timing model.  This
finding, and a similar anti-correlation between the time of arrivals
and the X-ray flux, may be evidence for a hot spot moving on the NS
surface ~\cite{pat10}. Similar conclusions were reached by other
authors to explain the timing noise observed in the timing residuals
of several AMXPs~\cite{pat09, har08, rig08, pap07}.  Some of these
observations can be understood in terms of the hot spot wandering
model described in Section~\ref{noise} although no final confirmation
of the models has yet been put forward. Despite many open problems
that need to be solved (such as why such correlations are not observed
in AMXPs like SAX J1808.4--3658) the moving hot spot model help us to
understand the AMXP pulse formation process.

Soon after the first AMXP discovery, it was realized that the lowest
energy (``soft'') photons that compose the pulsations arrive on
average later than the high energy (``hard'') photons~\cite{cui98}.
In SAX J1808.4--3658, hard photon arrival times tend to saturate at
some energy $E\gtrsim 10$ keV, with soft photons at about 2 keV
accumulating a lag of about 0.08 rotational cycles (or 200 $\mu$s).
Similar behaviour was soon discovered in other five AMXPs: XTE
J0929--314 (0.14 cycles~\cite{gal02}), XTE J1751--305 (0.06
cycles,~\cite{gie05}), XTE J1814--338 (0.016 cycles,~\cite{wat06}),
XTE J1807--294 (0.1 cycles, \cite{cho08}) and in the intermittent
pulsar HETE J1900.1--2455 (0.07 cycles,~\cite{gal07}).  In IGR
J00291+5934, soft lags were observed with hard photons leading by 0.06
cycles at 6-8 keV \cite{fal05b}.  Above this energy, harder photons
($E\gtrsim 8$ keV) start to gradually reduce their lead until at
energies of 30-100 keV there are no phase lags with respect to the
soft photons (at ${\sim}2$ keV). The origin of these lags is poorly
understood, and has been discussed in terms of the different angular
distribution (``fan'' and ``pencil'' beams) of a two spectral
component model~\cite{pou03} or in terms of Compton down-scattering of
hard X-ray photons from the cold disk plasma or the NS
surface~\cite{fal07}.  Detailed study of the soft lags in SAX
J1808.4--3658~\cite{har09a} suggests that they have a flux dependence,
with the lag being almost zero at high fluxes, increasing steadily at
lower fluxes and then decreasing again below a flux threshold
coincident with the onset of the rapid decay.  This may be linked with
the changing properties of the accretion disk as it transitions
towards the propeller phase.

\subsection{Pulse Shape Evolution}

Most AMXPs with strong harmonic content show pulse shape variability
that often correlates with the stage of the outburst. The pulse
profiles observed in \textit{different} outbursts of SAX J1808.4--3658
have characteristic shapes that can be associated with the specific
stage of the outburst (e.g., rise, peak, decay, etc..;
see~\cite{har08, ibr09, kaj11}). There is a very strong linear
anti-correlation between the fractional amplitude of the first
overtone and the X-ray flux (something very similar was also observed
in XTE J1807--294;~\cite{pat10}), suggesting that the secondary hot
spot becomes visible or more prominent as the accretion process
becomes less intense.  This may be explained if the inner disk radius
moves in or out with respect to the accretion rate, rendering the
secondary hot-spot visible ~\cite{har08, ibr09}.  If the inner disk
radius $r_{in}$ is assumed to vary at different outburst stages, then
the pulse profile changes of SAX J1808.4--3658 can be explained in
this way~\cite{pou09}. Variations in the disk-magnetosphere coupling
may also explain explain some of the pulse shape and phase variations
~\cite{kaj11}.  Calculation of the light-curve of accreting NSs in
global 3D-MHD simulations showed that if an octupolar magnetic field
dominates the field configuration, double peaked pulse profiles can be
reproduced~\cite{lon12}. Similar pulse shape variations were seen to
correlate with the outburst stage of Swift J1756.9--2508~\cite{pat10}
IGR J17511--3057~\cite{ibr09} and to some extent XTE J1807--294
\cite{pat10b}.  However, to date it is still unclear why pulse
profiles change abruptly and unpredictably in some sources during
certain outbursts but at other times have limited variability.

\section{Long Term Evolution and Pulse Formation Process}\label{sec:6}

If AMXP pulsations are observed in different outbursts one can follow the
long-term spin and orbital evolution.  To date only five AMXPs have been
monitored with high time resolution instruments in different
outbursts: SAX J1808.4--3658, IGR J00291+5934, XTE J1751--305, Swift
J1756.9--2508 and NGC6440 X-2 (although with relatively low S/N and
short outburst duration, it is difficult to constrain the long-term evolution of the latter two).

\subsection{Specific sources}

\runinhead{SAX J1808.4--3658}

\begin{figure}[t]
\sidecaption
\rotatebox{0}{\includegraphics[scale=.38]{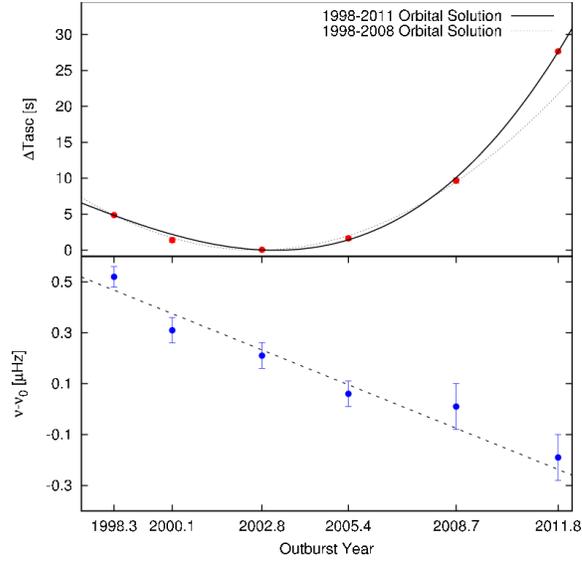}}
\caption{Orbital (top panel) and spin frequency evolution (bottom
  panel) of SAX J1808.4--3658 over a baseline of 13 years. The top
  panel shows the delays accumulated by the pulsar as it passes
  through the ascending node. The data were consistent until 2008 with
  a parabolic increase which is equivalent to a steady widening of the
  orbit (dotted line). The 2011 data revealed instead an acceleration
  of the orbit, as evidenced by the solid line in the top panel. The
  spin frequency has shown a steady decrease in between outbursts
  (bottom panel; $\nu_0$ is a frequency offset of $400.975210$ Hz),
  compatible with a spin down of the order of
  $-10^{-15}\rm\,Hz\,s^{-1}$. (Figure from~\cite{pat12}).}
\label{fig:1808}       
\end{figure}

The best constrained AMXP is SAX J1808.4--3658, for which secular spin
evolution and orbital period variation have now been measured over a
13 year baseline \cite{pat12,har08,har09}. As discussed previously,
accretion torques have a small effect on the spin of SAX J1808.4--3658
and have never been conclusively detected during any of the six
outbursts monitored by \textit{RXTE}. However, if one compares the
constant spin frequency measured in each outburst, SAX J1808.4--3658
is clearly spinning down at a constant rate (bottom panel in
Figure~\ref{fig:1808};~\cite{pat12}).  The stability of the spin-down
rate suggests that its origin is unrelated to propeller onset, since
spin-down would then depend on the amount of mass propelled and on the
duration of this phase, which can vary from outburst to outburst. Loss
of angular momentum via emission of gravitational waves was also
considered \cite{har08}, and is interesting since this might also
balance accretion torques, explaining the lack of measurable spin-up
during the outbursts. However, fine tuning is also required in this
case since the mechanism producing gravitational waves has to be
triggered for the right amount of time so as to guarantee a constant
spin-down between outbursts~\cite{pat12,has11,har09}.  The most likely
scenario appears to be a loss of angular momentum via magnetic-dipole
radiation. This is expected for any rapidly rotating NS with a
magnetic field, even if the NS does not turn on as a radio pulsar.
The measured spin-down is of the order of $-10^{15}\hzs$ and is
consistent with a polar magnetic field of $2\times10^{8}$
G~\cite{pat12, har08, har09}.

The long-term orbital evolution of SAX J1808.4--3658 is also
exciting. Theory states this should be driven by loss of angular
momentum via emission of gravitational waves from the tight binary
orbit.  For SAX J1808.4--3658, the resulting orbital evolution should
proceed on a timescale ${\sim} 10^9$ years~\cite{pac67}:
\begin{equation}
\tau_{\rm\,GW} = 0.01 \frac{\left(M_{NS} + M_2\right)^{1/3}}{M_{NS}\,M_{2}}P_{b}^{8/3} Gyr
\end{equation}
where $P_b$ is in hours and $M_{NS}$ and $M_2$ are the NS and donor
masses. The observed value, however, implies an expansion of the orbit
on a timescale of ${\sim}70$ Myr~\cite{har08, har09, dis08, bur09}.
Further observations in 2011 provided a long enough baseline to detect
an orbital period acceleration (top panel in
Figure~\ref{fig:1808};~\cite{pat12}).  SAX J1808.4--3658 may therefore
behave like the ``black widow'' binary millisecond pulsars observed at
radio frequencies~\cite{laz11} with a short-term exchange of angular
momentum between the donor star and the orbit. A similar suggestion
was also previously proposed for this AMXP~\cite{bur09, dis08},
invoking a ``hidden'' black widow scenario in which the AMXP switches
on as a radio pulsar during quiescence and ablates its donor.

\runinhead{IGR J00291+5934}

Long-term spin down has also been measured between the 2004 and 2008
outbursts of IGR J00291+5934.  The AMXP rotation behaves like a
saw-tooth, spinning up during the 2004 outburst then spinning down
until the 2008 outburst.  The spin-down rate is $\approx
-3\times10^{-15}\hzs$~\cite{pat10,har11,pap11}, comparable to that
observed in SAX J1808.4--3658. Although spin down between outbursts
has been observed only once and it is not possible (until the next
outburst) to verify whether it is constant, the most likely
explanation also involves magnetic dipole radiation spin-down.  If
this saw-tooth behaviour is a good indicator of the \textit{secular}
spin evolution of this AMXP, then the net spin frequency is
\textit{increasing} at $\approx +2.5\times10^{-15}\hzs$.  Indeed, the
spin frequency at the beginning of the 2004 outburst was less than
that observed in the 2008 outburst: the 2004 spin-up accelerated the
pulsar more than it subsequently decelerated in quiescence.  The
relatively small baseline did not permit any useful constraints on
orbital period evolution: this should be possible after the
observation of the next outburst.

\runinhead{XTE J1751--305}

Comparing the spin frequencies from 2002, 2005, 2007 and 2009 reveals
a secular spin down of $-5.5\times10^{-15}\hzs$~\cite{rig11}. This is
again consistent with magnetic dipole spin-down. If the spin-up
measured during the 2005 outburst reported in~\cite{pap08} is not
strongly affected by timing noise, then its behaviour resembles IGR
J00291+5934, with spin frequency increasing in outburst and decreasing
between outbursts in a saw-tooth. The inferred magnetic field is
${\sim}4\times10^{8}$ G, in line with the other two AMXPs constrained in
this way. The low quality of the data taken in the 2002 outburst does
not permit stringent constraints on orbital evolution.

\runinhead{Swift J1756.9-2508}

Observations during the outbursts in 2007 and 2009 give upper
limits of $|\nudot|\lesssim 2\times10^{-15}\hzs$ on the secular spin
evolution of this AMXP~\cite{pat10}. This requires a
magnetic field  $< 9\times10^{8}$ G. Magnetic field
estimates for all four of these AMXPs are given in Table~\ref{tab:B}.

\begin{table}
\scriptsize
\begin{center}
\rowcolors{3}{gray!10}{}
\begin{tabular}{lcccr}
\hline
\hline
Source & Spin-Down in Quiescence & Magnetic Field &  Secular Evolution & Reference\\
 & [$\rm\,Hz\,s^{-1}$] & [$10^{8}$ G] &\\
\hline
SAX J1808.4--3658 & $-10^{-15}$ & 1.5-2.5 & Spin-down &~\cite{pat12, har08,har09}\\
IGR J00291+5934 & $-3\times10^{-15}$ & 1.5-2.0 & Spin-up &~\cite{pat10,har11,pap11}\\
XTE J1751--305 & $-5.5\times10^{-15}$ & 4 & Spin-down &~\cite{rig11}\\
Swift J1756.9--2508 & $<|2|\times10^{-15}$ & $<9$ & ? &~\cite{pat09}
\end{tabular}\label{tab:B}
\caption{{\bf Secular Evolution and Inferred Magnetic Fields in
    AMXPs}. The magnetic field is inferred from the spin down observed
  in quiescence, and refers to field at the poles of the NS for a pure
  dipolar configuration, $R=10$ km and $M=1.4\msun$.}
\end{center}
\end{table}

\subsection{The Maximum Spin Frequency of Neutron Stars}

None of the AMXPs discussed above has a spin rate that \textit{increases} on long timescales,
except for IGR J00291+5934. Even this source has a \textit{net}
acceleration so small that its spin frequency will change
significantly only on timescales of several billion years.  This raises the
question on whether this behavior is the norm for AMXPs.

\begin{figure}[t]
\sidecaption
\rotatebox{-90}{\includegraphics[scale=.50]{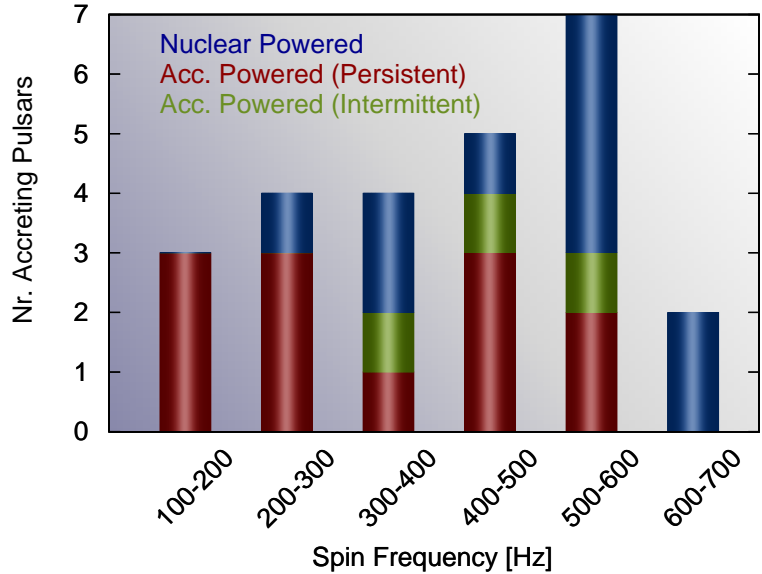}}
\caption{Histogram of the spin frequency of AMXPs (both intermittent
and persistent pulsators) and NXPs. The
  histogram is empty for frequencies larger than about 700 Hz. This is
  similar to what is found in radio pulsar data, where the sample size
  is considerably larger (with hundreds of millisecond radio
  pulsars).}
\label{fig:histo}       
\end{figure}

For all realistic EoS of ultra-dense matter, NSs are stable at spin
frequencies well in excess of 1000 Hz.  The break-up frequency is well
approximated by
\begin{equation}
\nu_{\rm\,max} = 1230 \left(\frac{M_{NS}}{1.4 M_{\odot}}\right)^{1/2}\left(\frac{R_{NS}}{10\rm\,km}\right)^{-3/2}\rm\,Hz
\end{equation}
(where $M_{NS}$ and $R_{NS}$ refer to the non-rotating mass and radius
of the NS under consideration, expression valid for arbitrary NS mass
and EoS as long as the mass is not too close to the maximum permitted
for that EoS~\cite{lat04, lat07}).  However the distribution of spin
frequencies of the ensemble of AMXPs and NXPs (nuclear powered X-ray
pulsars, see Introduction and Section 7) has an abrupt cutoff at about
730 Hz. This was first noticed in 2003~\cite{cha03} and has been
confirmed in later works with larger sample size~\cite{cha08, pat10b}.
Figure~\ref{fig:histo} shows the current distribution of spin
frequencies for the 15 AMXPs and the 10 NXPs known (see Table 1 in
\cite{Watts12} for a complete list of NXPs). So far no AMXP or radio
millisecond pulsar has been found above this cutoff (the fastest radio
millisecond pulsar has a spin of 716 Hz~\cite{hes06}). So not only
there are no pulsars with $\nu_s\gtrsim700$ Hz, but also at least four
AMXPs that should be accelerating in response to accretion are instead
either decelerating or spinning up very slowly, on timescales of
billions of years.

Gravitational waves have been invoked to explain this cutoff, but as
shown recently~\cite{har11, pap11, has11}, this cannot be the
explanation for all AMXPs as it would require substantial fine-tuning
to explain the X-ray timing observations. The cutoff might instead be
related to the magnetic field evolution of NSs~\cite{pat12a}.
However, further investigation is needed to assess this fascinating
question in a robust way since it is still unclear how the magnetic
field of NSs evolves in response to accretion~\cite{rom90, wij97,
  wij99, kon02, kon04, bej11}.

\subsection{Why do most Low Mass X-ray Binaries not pulsate?}\label{notpulsate}

Even after 15 years of high time resolution X-ray observations, only a
few LMXBs have shown millisecond pulsations. Several LMXBs have
pulsations at long periods (Table~\ref{tab:lmxbs}) but the number of
pulsating systems is still small compared to the entire LMXB
population.  Several mechanisms have been proposed to solve this
problem, including burial of the magnetic field by
accretion~\cite{bis74, rom90, cum01}, smearing of the pulsations by an
optically thick corona~\cite{bra87, tit02, tit07}, smearing of
pulsations due to gravitational light bending~\cite{oze09, woo88},
alignment of the NS magnetic and rotational axes~\cite{rud91, lam09}
and onset of MHD instabilities at the disk/magnetospheric boundary
\cite{rom08}. None of these models have yet been confirmed, although
several new findings have helped to test their predictions.

\runinhead{Magnetic Field Screening.}

In this model, fresh unmagnetized accreted plasma diamagnetically
screens the NS magnetic field as it is deposited on the
surface~\cite{bis74, rom90, cum01}.  The field then re-emerges via
Ohmic diffusion and competition between the Ohmic diffusion and
accretion timescale sets the behaviour of the magnetic field. If the
accretion timescale is shorter than the Ohmic diffusion timescale, the
magnetic field is buried deep in the ocean and outer crust.  The two
timescales can be defined as:
\begin{eqnarray}
\tau_{\rm\,acc} &=& \frac{y}{\dot{m}}\\ \tau_{\rm\,diff} &=& \frac{H^2}{\eta}
\end{eqnarray}
where $y=\rho dz$ is the column density, $z$ is the direction
orthogonal to the surface in the plane parallel geometry
approximation, $\dot{m}$ is the local mass accretion rate per unit
area (in units of $\rm\,g\,cm^{-2}\,s^{-1}$), $H=\frac{d lnP}{dz}$ is
the pressure (defined as $P$) scale-height, and $\eta$ is the magnetic
diffusivity coefficient. This model predicts that LMXBs with very low
mass accretion rates should be AMXPs, and that the most luminous LMXBs
should not pulsate.  This seems to be the case, since persistently
pulsating AMXPs are indeed all faint LMXBs. However, the discovery of
intermittent pulsations as short as ${\sim}100$ seconds in Aql X-1
poses severe problems for the model since it is unclear how a buried B
field could emerge for this brief episode. Another interesting
observational result comes from HETE J1900.1--2455, which, with a mean
bolometric luminosity of $4.4\times10^{36}\ergs$, has a mean mass
accretion rate $\mdot\,\approx8\times10^{-10}\msun\rm\,yr^{-1}$ for an
assumed accretion efficiency of 20\%. This accretion rate is about
twice that of most AMXPs, but is smaller than the bolometric
luminosity of SAX J1808.4--3658 during observations taken close to the
peak of the 2008 outburst~\cite{kaj11}. This might indicate that the
screening timescale is long compared to the outburst duration of most
AMXPs.  In HETE J1900.1--2455 observational evidence for magnetic
field screening has been found~\cite{pat12c}, with the magnetic field
possibly reducing by orders of magnitude on a timescale of about 100
days during which channeled accretion becomes less and less efficient
until it stops working. If confirmed this indicates that indeed most
LMXBs might be not pulsating because of the lack of an extended
magnetosphere. Recent 2-D and 3-D numerical simulations have started
to address this problem and results seems to suggest that
submergence of the magnetic field is possible at least when the
accretion rate is high (\cite{ber13} but see also \cite{muk13,pay07,pay04}
for different results). Despite still open problems, the screening model remains
therefore one of the most promising, although further efforts
are necessary to formulate new sophisticated predictions and tests.

\runinhead{Smearing from an Optically Thick Corona.}

This model suggests that X-ray emission from AMXPs is formed primarily
via Comptonization processes in a relatively optically thin medium.
If the Compton cloud is thick, then pulsations can be washed out.
This scenario has been criticized on the grounds that spectral
analyses of most LMXBs show small optical depths, well below the limit
required to smear out pulsations~\cite{gog07}, which should therefore
be visible (but see~\cite{tit07}).  It is not clear why in this model
the intermittent AMXP SAX J1748.9--2021 shows pulsations that appear
and disappear on timescales of a few hundred seconds~\cite{alt08,
  pat09}, uncorrelated with spectral changes.  However it provides a
reasonable explanation for the time lags and pulse energy dependence,
especially for the pulses of IGR J00291+5934 which have a turnover at
7 keV that is otherwise difficult to explain~\cite{fal05b, fal07}.

\runinhead{Gravitational Light Bending.}

For large stellar compactness ($M/R$), gravitational light bending can
distort the path of X-ray photons up to the point where pulsations are
strongly reduced~\cite{woo88, bel02, oze09}. Persistent sources may
therefore be less likely to show pulsations because they accrete more
mass than transient systems.  However, the total amount of mass that a
NS has accreted does not depend on the current mass accretion rate and
on the persistence of the X-ray flux, but on the evolutionary stage of
the donor star and on the average mass transfer rate from the donor
towards the accretor.  This model also requires that the NSs in most
LMXBs are more massive that those in AMXPs.  Although possible, this
is difficult to reconcile with the fact that most AMXPs have probably
accreted for a long time (see for example~\cite{nel03}).  The model
also requires some special configuration of the accretion column,
which existed for only a very short time and has never recurred, to
explain the short pulsed episode of Aql X-1. Finally, the light
bending effect never suppresses completely oscillation amplitudes
since the Doppler shift due to the rapid NS rotation introduces
anisotropies in the emission pattern and thus pulsations~\cite{vii04,
  pou06b}.

\runinhead{Magnetic and Rotational Axes alignment.}

The possibility that the magnetic poles of NS migrate was first
suggested by Ruderman in 1991~\cite{rud91}. Vortex/fluxoid coupling in
the NS interior links the spin and magnetic axes.  Once the NS is
spun-up by accretion, crustal shear stress breaks the crust into
``plates'' that drift towards the rotational poles. The phenomenology
of an AMXP with a nearly aligned magnetic and rotational axis was
developed recently~\cite{lam09} and explains fairly well the observed
correlations between pulse amplitude, pulse phase and X-ray flux. The
model can explain intermittency~\cite{lam09b}, but implies that the
persistent AMXPs should also show occasional ``oscillation dropouts'',
a sudden and brief disappearance of pulsations due to sporadic
alignment of the hot spot and rotational axis.  Such dropouts have not
been observed, although the timescale over which this might happen may
be too short compared to timescale over which the pulsations are
measured (i.e., hundreds of seconds).  The model also predicts
correlations between excess background noise produced by the hot spot
motion which is anti-correlated with the pulse amplitude. This
prediction could in principle be easily verified in the future with
observational tests.

\runinhead{MHD Instabilities.}

Accretion onto a magnetized NS can proceed via channeled accretion but
also via the Schwarzschild-Kruskal instability (the magnetic version
of the Rayleigh-Taylor instability) in the disk equatorial plane
(\cite{aro76, els77}). This latter mode of accretion has now been
observed in 3D-MHD simulations (\cite{kul08}) and predicts a
decoherence of the X-ray signal at high accretion rates, when several
``tongues'' of plasma penetrate the magnetosphere and impact the NS
surface at random positions.  This model explains why bright LMXBs do
not pulsate, and might explain why intermittent sources are on average
brighter than persistent AMXPs.  However, the intermittent AMXP SAX
J1748.9--2021 has shown pulsations that appear and disappear in a
broad range of accretion rates~\cite{pat09} (rather than at a sharp
threshold).  It is also hard to explain why tongues never develop in
any other AMXP when the outbursts reach peak luminosity, during which
the accretion rates are comparable or higher than observed in
intermittent AMXPs.

\section{Thermonuclear Bursts}\label{sec:7}

\begin{table}
\scriptsize
\begin{center}
\rowcolors{4}{gray!10}{}
\begin{tabular}{lllr}
\hline \hline Source & Burst osc. & Phase locking &
References\\

                         & & & \\ \hline \textbf{Accreting Millisecond
  X-Ray Pulsars}\\ \hline SAX J1808.4--3658 & Yes & No & ~\cite{int98,
  int01, cha03, Bhattacharyya06b, gal06, Bhattacharyya07b, galcat08}
\\ XTE J1814--338 & Yes & Yes & ~\cite{str03, Bhattacharyya05,
  Watts05, Watts06, wat08, galcat08} \\
\smallskip
Aql X--1 & Yes & No & ~\cite{Koyama81, Czerny87, Lewin93, zha98,
  Muno01, Muno04, galcat08} \\ SAX J1748.9--2021 & No & &
~\cite{int99, Kaaret03, alt08, galcat08} \\ HETE J1900.1--2455 & Yes &
No & ~\cite{Kaaret06, suz07, gal07, galcat08, wat09b} \\
\smallskip
IGR J17511--3057 & Yes & No & ~\cite{alt10d, boz10, Falanga11, rig11c}
\\ IGR J17498--2921 & Yes & Unknown & ~\cite{lin11c, cha12} \\ Swift
J1749.4--2807 & No & & ~\cite{wij09, fer11} \\  IGR
J18245--2452 & Yes & Possible & ~\cite{pat13a, pap13a,rig13} \\\hline \textbf{Mildly
  Recycled X-ray Pulsar}&&&\\ \hline IGR J17480-2466 & Yes & Yes
&~\cite{cav11, mot11, chakraborty11b, chakraborty11, lin11b, lin12}
\\ \hline \hline
\end{tabular}\\
\caption{{\bf Burst oscillation properties for the nine AMXPs that
    have shown thermonuclear bursts}. The table indicates whether a
  source shows burst oscillations and whether their phases match and
  follow the evolution of the accretion-powered pulses observed before
  and after the bursts (``Phase Locking'').}
\end{center}
\label{tab:bursts}
\end{table}

Type I X-ray bursts are thermonuclear explosions triggered by unstable
burning of hydrogen or helium on the layers that build up on the
surfaces of accreting NSs. The basic cause is an imbalance between
nuclear heating and radiative cooling, which leads to the runaway of
temperature-dependent thermonuclear reactions \cite{Lewin93}.  There
are however many aspects of burst physics that remain puzzling, such
as discrepancies between predicted and observed recurrence times, and
the phenomenon of burst oscillations \cite{str06, Watts12}.  The
latter are high frequency oscillations (11-600 Hz) often detected in
the X-ray flux produced by thermonuclear bursts~\cite{galcat08}, and
whose cause remains unknown. AMXPs can offer particular insight into
burst physics and the burst oscillation mechanism, for two reasons.
Firstly, they are the only sources where we have an independent
measurement of the stellar spin, allowing us to measure the dependence
on rotation. Secondly, they are the only sources where we can make a
reasonable assessment of the effects of the magnetic field, and the
role of uneven fuel distribution (due to magnetic channeling of
material onto the poles of the star).

Table 5 summarizes the AMXPs that have exhibited
bursts, and which of these have shown burst oscillations. Bursts are
expected and observed to occur only within a range of accretion rates
from ${\sim} 2-30$\% of the Eddington rate~\cite{Fujimoto81,
  Cornelisse03, Narayan03}, and in this regard the occurrence of
bursts in the AMXP population accords with the behaviour seen in the
non-pulsing bursters.  The sources that do not show bursts are the
five ultracompact sources (which are thought to have rather low
accretion rates~\cite{str06}, likely below the minimum required for
regular bursting ~\cite{int07}) and IGR J00291+5934, which is also
thought to have a rather low accretion rate~\cite{gal05}. None of the
AMXPs has yet shown a super-burst, longer duration bursts thought to
be due to unstable carbon burning~\cite{Kuulkers04}.

\runinhead{Burst properties.} Bursting behaviour is known to be
extremely variable, and the bursts of the AMXPs are no exception to
this rule.  To date there has not been a comprehensive study assessing
whether the bursts from the AMXPs have properties that differ in any
statistically significant way from the non-pulsing sources.  However
just as for the non-pulsing bursters, bursts from the AMXPs show
variable recurrence times, peak fluxes and fluences, and durations,
even when accretion rate seems to be relatively stable.  There are
nonetheless some features of AMXP bursts that are worthy of note, in
particular when they are compared with bursting sources with higher
magnetic fields like the mildly recycled pulsar IGR J17480--2466.

\subruninhead{Magnetic fields and bursting.}  Bursts from AMXPs
confirm that magnetic fields ${\sim} 10^8-10^{10}$ G (Table
\ref{tab:lmxbs}) are no impediment to thermonuclear bursting.

\subruninhead{Burst recurrence times.} 

Many neutron star low mass X-ray binaries show burst
recurrence times of less than an hour, something that is problematic
for theoretical models \cite{Keek10}.  The only AMXP in this short
recurrence time group is the highly intermittent source Aql X--1. By
contrast the mildly recycled pulsar IGR J17480--2466 has the shortest
recurrence time between thermonuclear bursts yet recorded, at 3.3
minutes~\cite{mot11}, a possible sign of confined burning.
  
  \subruninhead{Rotation and bursting.} 

The AMXPs, which have independent estimates of spin rate, confirm that
bursting can occur for spin rates of up to a few hundred Hz (although
the fastest rotating burster, 4U 1608--522, is a burst oscillation
source not an accretion-powered pulsar, so the spin measure is more
indirect ~\cite{Hartman03}).  The mildly-recycled
accreting pulsar IGR J17480--2466 is instead the bursting source with burst
oscillations with the lowest measured rotation rate, at 11 Hz.  For a
star rotating at ${\sim} 10$ Hz, the Coriolis force is not dynamically
relevant~\cite{cav11,Watts12}, indicating that Coriolis force induced
confinement of the igniting patch is not essential to the development
of X-ray bursts~\cite{spi02} or to very short burst recurrence times
(see previous point).
  
\subruninhead{Cooling as a signature of thermonuclear bursts.}  

Cooling during the tails of bursts (inferred from blackbody fits to
the burst spectra) has long been regarded as a signature trait of
thermonuclear Type I (rather than accretion-powered Type II) bursts.
Detailed studies of the bursts from the mildly recycled pulsar IGR
J17480--2466 have now shown that cooling is a sufficient but not
necessary condition for the identification of thermonuclear bursts:
under certain circumstances, particularly at high accretion rates, the
cooling signature may not be readily detectable~\cite{chakraborty11b,
  chakraborty11, lin11b, lin12}.  Bursts from the AMXPs do however
show cooling.
  
\subruninhead{Marginally stable burning.} 

  The mildly recycled pulsar IGR J17480--2466 shows mHz QPOs that are
  interpreted as marginally stable burning of H and He on the NS
  surface~\cite{Heger07,lin12}. The mHz QPOs are observed when the
  accretion rate rises and the source transitions from unstable
  burning (X-ray bursts) to stable nuclear burning (no bursts). The
  only AMXP to have shown mHz QPOs is the highest accretion rate
  source, the intermittent pulsar Aql X--1 \cite{rev01}.

\subruninhead{Pure helium bursts.} 

SAX J1808.4--3658 has very bright, short bursts that are thought to be
the only secure example of bursting occurring in a layer of pure
helium, for a source where the accreted fuel is a mix of hydrogen and
helium~\cite{gal06}.  The only other sources that are thought to have
pure helium bursts are ultra-compact sources where the donor star is
too small for there to be significant hydrogen in the accreted
material.  Pure helium bursting is thought to occur in mixed fuel
sources only in a very narrow range of accretion rates where hydrogen
can burn stably to form helium before the helium ignites
unstably~\cite{Fujimoto81}.
  
\subruninhead{Bursting and intermittent accretion-powered pulsations.}
  Two of the intermittent AMXPs, HETE J1900.1--2455 and SAX
  J1748.9--2021, show intermittent pulsations with variable amplitudes
  at times when the source is also bursting~\cite{gal07, alt08}.
  There does not appear to be any clear-cut causal relationship
  between the two phenomena (and the third intermittent AMXP, Aql X-1,
  does not show bursts at the time of its one intermittent pulsation
  episode,~\cite{cas08}).  Nonetheless there has been speculation in
  the literature as to whether the two might be related.  Radiation
  from the burst might perhaps lead to a temporary disruption of the
  inner edge of the disk, perturbing the flow of material onto the
  magnetic poles and moving the accretion footprint to a location more
  favorable for the formation of pulsations~\cite{lam09b}.

\runinhead{Burst oscillations} 

Where the AMXPs have had the biggest impact is in our understanding of
burst oscillations (see~\cite{Watts12} for a more comprehensive review
of this topic).  Burst oscillations were first observed in 1996, from
the non-pulsar 4U 1728--34~\cite{Strohmayer96}.  The oscillations
manifested as a coherent signal at about 363 Hz, with an upwards drift
in frequency of ${\sim} 1$ Hz as the bursts progressed, towards an
asymptotic maximum in the tails.  The fact that the same frequency was
seen in multiple bursts from the same source, and the stability of the
asymptotic maximum frequency, suggested a link to a stable clock such
as the stellar spin rate~\cite{Strohmayer98b}.  Firm identification of
burst oscillation frequency with the spin frequency however had to
wait until the first robust detection of burst oscillations from an
AMXP~\cite{cha03}.  Although the two frequencies appear to be very
close (within a few Hz), however, the separation does differ markedly
for different AMXPs.  The AMXP XTE J1814--338 and the mildly recycled
pulsar IGR J17480--2466 have burst oscillation frequencies that agree
with the spin frequency very closely (within the error bars, ${\sim}
10^{-8}$ Hz for XTE J1814--338~\cite{str03, Watts05, wat08},
${\sim}10^{-4}$ Hz for IGR J17480--2466~\cite{cav11}).  SAX J1808.4--3658
and IGR J17511--3057, by contrast, have burst oscillations that show a
rapid increase in frequency in the burst rise, slightly overshooting
the spin frequency, and then stabilizing to within ${\sim} 10$ mHz of
the spin frequency in the tails~\cite{cha03, alt10d}.  The
intermittent pulsars Aql X--1 and HETE J1900.1--2455 show burst
oscillations with slow drifts throughout the bursts, more similar to
those of the non-pulsars, and in these cases the asymptotic maximum
frequency is ${\sim} 1$ Hz below the spin frequency~\cite{zha98, mun02,
  cas08, wat09b}.  Any model for the burst oscillation mechanism must
be able to explain this diversity in frequency drift and the small
variations in offset from the spin frequency.

There are interesting differences in the properties of the burst
oscillations from the pulsars as compared to those of the non-pulsars.
For the non-pulsing bursters, burst oscillations tend only to be seen
when the source is in the soft (high accretion rate) state
\cite{galcat08}.  The two intermittent pulsars conform to this rule:
indeed HETE J1900.1--2455 has shown burst oscillations in only one
burst, when the source entered an unusually soft state~\cite{wat09b}.
By contrast five of the six persistent pulsars with bursts have had
burst oscillations in all of their bursts, even though these sources
tend to be in the hard (low accretion rate) state.  For the fifth, IGR
J17498--2921, the data are of much poorer quality: oscillations are
seen only in the brightest bursts but the upper limits on the presence
of detections in the weaker bursts are comparable to the amplitudes
detected in the brighter bursts~\cite{cha12}.  The duration of burst
oscillation trains also appears to differ.  For SAX J1808.4--3658, XTE
1814--338 and the mildly recycled pulsar IGR J17480--2466, burst
oscillations persist throughout the bursts (except during episodes of
Photospheric Radius Expansion in the bright bursts from SAX
1808.4--3658).  For the non-pulsars, although burst oscillations are
sometimes detected throughout bursts they are more commonly detected
in the tails~\cite{galcat08}, and this is also the case for the
intermittent accretion-powered pulsars.  In this regard the persistent
pulsar IGR J17511--3057 seems to be a transitional object: at lower
accretion rates oscillations are detectable throughout bursts, but as
accretion rate rises the burst oscillation signal vanishes from the
burst rise~\cite{alt10d}.  Burst oscillations from the pulsars have
higher harmonic content than burst oscillations from the intermittent
pulsars and non-pulsars \cite{mun02b, cha03, str03, Watts05, wat09b,
  alt10d, cav11}.  They also have rather different amplitude-energy
relations.  Burst oscillations from the persistent pulsars show the
same amplitude-energy relationship as their accretion-powered
pulsations, irrespective of whether the latter rise or fall with
energy but with the latter being more common (\cite{cha03, str03,
  wat09b, alt10d, cav11, papitto, cha12} and see also
Section~\ref{sec:amp}). The intermittent pulsars and non-pulsars, by
contrast, have burst oscillation amplitudes that rise with energy
\cite{MunoB, wat09}.  The cause of these various differences in burst
oscillation properties between the pulsars and non-pulsars remains as
yet unclear, but the most obvious hypothesis is that the differences
are due to the effects of a dynamically important magnetic field.

One can also use the pulsars to compare the properties of the
accretion-powered pulsations (where the peak of the emission is
presumed to be centered on the magnetic poles) and their burst
oscillations.  Burst oscillation amplitudes vary substantially from
source to source, but are in most cases within a few percent of (and
most lower than) the amplitude of the accretion-powered pulsations at
the time of the burst~\cite{cha03, str03, Watts05, wat09b, alt10d,
  cav11, papitto}.  The transitional AMXP/radio pulsar IGR
J18245--2452 is an exception: for the one burst detected, the burst
oscillation amplitude was much higher than the amplitude of
accretion-powered pulsations immediately before the
burst~\cite{pap13a}.

Pulsar burst oscillations show no statistically
significant phase lags as a function of energy \cite{MunoB, Watts06,
  wat09b}, unlike accretion-powered pulsations which show soft lags
(\cite{cui98, Gierlinski02, gal02, Kirsch04, Gierlinski, pap10}; see
also Section~\ref{sec:amp}). Of particular interest is the strong
phase-locking between accretion-powered pulsations and burst
oscillations seen in XTE J1814-338 and IGR J17480-2466~\cite{wat08,
  cav11}, with very small phase offset ( $<3^{\circ}$ and
$\lesssim10^{\circ}$ respectively), implying that the hot spots
responsible for the two sets of pulsations are longitudinally
coincident in these two sources.

So what do these results tell us about the burst oscillation mechanism
? Although the mechanism is not yet understood, substantial progress
has been made and the accretion-powered pulsars have played a key
role. Their primary contribution, of course, has been to highlight the
very close relationship between the burst oscillation frequency and
the spin frequency of the star.  This fact underpins the two main
classes of model: hot spot models and global mode models.  Small
temperature variations in the surface layers mean that ignition is
expected to begin at a point, with a flame front then spreading out
across the star~\cite{Shara82}.  This flame may then either stall,
confining the burning to a small region (the hot spot model) or excite
large-scale waves in the surface layers (global mode models).  The
resulting temperature asymmetry gives rise to the burst oscillations.
In hot spot models, the relationship with the spin rate is
straightforward since the hot spot should be near stationary in the
rotating frame of the star~\cite{Strohmayer96}.  The main open
theoretical question for hot spot models is what might cause the flame
to stall, with both magnetic and Coriolis forces under consideration
\cite{bro98, spi02, cav11}.  For global mode models, the requirement
that burst oscillation frequency be very close to spin frequency puts
very stringent restrictions on the types of surface modes that might
be responsible~\cite{Heyl04, Piro05, Heng09}.  Efforts are now ongoing
to develop both classes of model to determine whether they could match
the rest of the observed properties~\cite{Watts12}.  The differences
between the burst oscillations from the pulsars and the non-pulsars
are informative, and suggest a role for the magnetic field.  What is
still unclear, however, is whether we are seeing one mechanism with a
continuum of properties set by stellar parameters such as the magnetic
field - or whether two different mechanisms may be required.  The
apparent gradual changes in burst oscillation properties from the
non-pulsars and intermittent pulsars (where magnetic field is likely
to be weak) to the persistent pulsars (with stronger fields) had until
recently favoured the former. However new results from the pulsars
point increasingly towards the possibility that we are indeed seeing
two different mechanisms.  The development of burst oscillations in
the slowly rotating mildly recycled pulsar IGR J17480-2466 is
particularly hard to understand in the context of global modes or
Coriolis force induced hot spots~\cite{cav11}. Magnetic confinement of
the flame front is at present the only plausible explanation for this
source, and would also explain the extraordinary phase-locking between
accretion-powered pulsations and burst oscillations.  This mechanism
might also operate in the other strong magnetic field source with
phase-locked burst oscillations, the AMXP XTE J1814-338~\cite{wat08,
  cav11}.  However for the other sources, with weaker magnetic fields,
magnetic confinement is unlikely to be effective.  For these sources
global mode models remain a good possibility, with the differences in
the pulsar burst oscillations being perhaps due to magnetic
modifications to the mode structure.

\section{Aperiodic Variability and kHz QPOs}\label{sec:8}

The study of aperiodic signals in power spectra of AMXPs reveals a
rich phenomenology which has helped us to understand the physics of
accretion disks and compact objects. All AMXPs are ``atoll sources'',
so-called because of their pattern in the X-ray colour-colour
diagram. Atoll sources are relatively low/intermediate-luminosity
LMXBs and are believed to possess low magnetic fields compared to the
brightest LMXBs (the ``Z sources''). The aperiodic variability of
AMXPs includes phenomena at very high frequencies (the kHz QPOs) but
also at lower frequency, such as the strong 1 Hz modulation observed
in SAX J1808.4--3658 and NGC6440 X-2 (Sections~\ref{sec:1808} and
\ref{sec:6440}). Here we highlight the main aperiodic phenomena
peculiar to AMXPs and refer to Chapter~1 of this book for a more
detailed discussion.

\begin{figure}[t]
\sidecaption
\includegraphics[scale=.24]{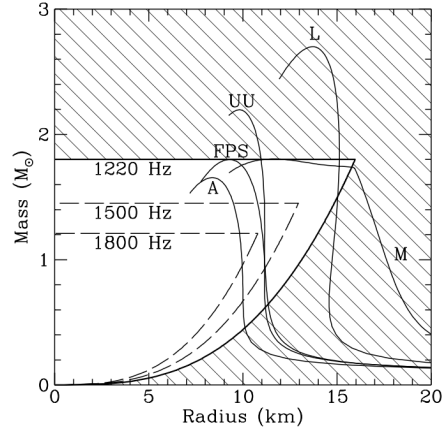}
\caption{Constraints on the Mass-Radius relation of NSs from kHz
  QPOs. The solid curves refer to a sample of different EoS models
  (for non-rotating NSs), and the thick solid and dashed lines enclose
  regions of the M-R diagram compatible with the observation of kHz
  QPOs at a certain Keplerian frequency (image from ~\cite{mil98b}).}
\label{fig:qpo}       
\end{figure}

The fastest variability in LMXBs is observed as kHz QPOs, which appear
at frequencies of up to ${\sim}1300$ Hz. Sometimes two kHz QPOs are
observed simultaneously as ``twins'', with an ``upper'' QPO at
frequency $\nu_u$ and a ``lower'' QPO at frequency $\nu_l$. Such fast
variability is observed only in LMXBs with NS accretors, and is
considered to be a signature of the presence of a NS in the absence of
pulsations or thermonuclear bursts. All models that try to explain kHz
QPOs associate at least one of the frequencies with the orbital motion
of plasma in the accretion disk. Such high frequency variability is
expected given the short dynamical timescale (0.1-1 ms) characterizing
the motion of matter near NSs.  Such rapid variability is never
observed in black hole LMXBs\footnote{Fast variability appears in
  black hole LMXBs below ${\sim}500$ Hz as ``high frequency QPOs''.},
possibly because they have larger mass and hence a larger innermost
stable circular orbit (ISCO). The ISCO is a limiting radius (at
$6GM/c^2$ for non-rotating NSs) set by General Relativity that defines
the smallest stable orbit in an accretion disk. The frequency of
kHz QPOs is close to the value expected for stable Keplerian orbits
around a NS, and can therefore be used to place constraints on the
mass and radius of the NS (and hence the EoS of ultra-dense
matter). The size of the orbit where kHz QPOs are produced must be
larger than both NS radius and ISCO, both of which depend on NS mass
(Figure~\ref{fig:qpo}).

The four AMXPs with twin kHz QPOs (SAX J1808.4--3658, XTE J1807--294,
IGR J17511--305 and Aql X-1; Section~\ref{sec:3}) have been used to
test kHz QPO formation models (e.g., the rejection of the
beat-frequency model; Section~\ref{sec:1808}) and have illuminated the
relation with the NS spin. Taking into account also the NXPs with twin
kHz QPOs, it is clear that the relation between kHz QPO separation
$\Delta\nu=\nu_u-\nu_l$ and spin frequency $\nu_s$ is not as simple as
originally thought: $\Delta\nu/\nu_s$ is not always equal to 1 or
0.5. $\Delta\nu$ and $\nu_s$ may even be unrelated: instead we see an
average $\langle\Delta\nu\rangle\approx 300$ Hz for all sources
\textit{except} AMXPs~\cite{men07, yin07}.  \textit{Only} for the
AMXPs, where the magnetic field is strong enough to channel accretion,
is the ratio $\Delta\nu/\nu_s$ either 1 or 0.5.  Some predictions of
the sonic-point spin resonance and relativistic-resonance
models~\cite{lam03,klu05} are consistent with the AMXP results. In
sonic-point spin resonance models, the magnetic and radiation fields
rotating with the star excite a vertical motion of the disk gas at the
``spin-resonance'' radius $r_{sr}$. There, the vertical epicyclic
frequency equals the difference between the orbital frequency and the
spin.  Depending on the ``clumpiness'' of the flow at $r_{rs}$, the
X-ray flux exhibits $\Delta\nu$ equal to either $\nu_s$ (smooth flow)
or $0.5\nu_s$ (clumpy flow).  In relativistic resonance models, a
non-linear 1:2 or 1:3 resonance between orbital and radial epicyclic
motion emerges as a consequence of the deviation of the strong
gravitational potential from the Newtonian $1/r$.  A third model
suggests that kHz QPOs are related to the three general relativistic
epicyclic frequencies~\cite{ste99} or to the precession of frame
dragging (Lense-Thirring precession~\cite{ste98}). So far none of
these models satisfactorily accounts for all of the rich phenomenology
of kHz QPOs.

Determining the origin of QPOs, in particular kHz QPOs is of
fundamental importance as this is a phenomenon that is common in
LMXBs.  If a clear relation with the NS mass and/or radius is
confirmed, it has the potential to place severe constraints on the EoS
of ultra-dense matter. This field has however seen few advancements in
the past few years, at least on the theoretical side.  New attempts
are now using numerical simulations to interpret the observations and
test some of the original models~\cite{dex11, ing11, ing12, bac10}.
Most of these simulations use black hole accretors (but
see~\cite{bac10}) which do not have kHz QPOs, but the results are
promising in terms of understanding the generation of variability in
LMXBs as a whole.

\section{Open Problems and Final Remarks}\label{sec:9}

Since the first discovery in 1998, the AMXPs have provided an
incredibly body of observational data allowing us to understand NSs
and their evolution.  They have allowed the study of extreme phenomena
on the surface of the NSs, in the strong gravity regime at
relativistic rotational velocities, and have helped us to understand
how radio pulsars form and evolve within the framework of the
\textit{recycling scenario}.

Proposed X-ray missions like the Indian \textit{ASTROSAT}, the NASA
\textit{AXTAR} satellite and the ESA missions \textit{LOFT} and
\textit{Athena+} will guarantee huge advances in our knowledge of
these systems, building on the extraordinary results from
\textit{RXTE}, \textit{XMM-Newton}, \textit{Chandra} and
\textit{Swift}.  Long-term monitoring of future AMXP outbursts will
allow measurement of long-term spin and orbital behaviour, something
that has at present only been done for four systems.  Increased
collecting area will increase sensitivity to pulsations by orders of
magnitude, leading (we hope) to the discovery of sub-millisecond
pulsars or strengthening the 730 Hz spin distribution cutoff. In the
former case this would place very tight constraints on the EoS of
ultra-dense matter.  Modeling of high signal to noise pulse profiles
will reveal the subtle features imparted by strong gravity, enabling
us to measure the NS compactness.  We will be able to test models for
kHz QPOs and burst oscillations with sufficient precision that we can
finally determine the underpinning mechanism.

Of the many open problems, there are several that stand out. One big
mystery is why do so very few LMXBs show pulsations?  If magnetic
field evolution and decay is involved, we may finally be able to
understand the origin and the evolution of nature's strongest magnetic
fields. What happens when accretion takes place during an outburst?
Why we do not detect the effect of accretion torques in some AMXPs,
contrary to expectations ? Are gravitational waves involved as a
braking mechanism that counteracts the effect of accretion or are we
again missing some fundamental ingredient?  When do accreting pulsar
switch on as radio pulsars ? Are all AMXPs turning on as radio
millisecond pulsars during quiescence ? What is the relation between
AMXPs and the radio pulsar ``black-widows'' and ``red-backs'' ?  We
look forward to discovering the answers to these questions in the
years ahead.

\begin{acknowledgement}
  We would like to thank C. Heinke, J. Poutanen, R. Lovelace,
  M. Linares, T. Tauris, D. Altamirano, B. Haskell, P. D'Avanzo,
  D. Bhattacharya, J. Hessels and N. Masetti for providing useful
  comments and suggestions. We acknowledge support from an NWO Veni
  (AP) \& Vidi (AP, AW) fellowships.
\end{acknowledgement}
%
\addcontentsline{toc}{section}{Appendix}
%
%
%

\end{document}